\documentclass[twocolumn,aps,pra,amsmath,amssymb,superscriptaddress]{revtex4-1}

\usepackage[usenames]{color}
\usepackage{hyperref}
\usepackage{graphicx}

\def\<{\langle}
\def\>{\rangle}

\begin{document}

\title{Bunching, clustering, and the buildup of few-body correlations in a quenched unitary Bose gas}
%\title{Bunching and clustering:  Correlation buildup in a quenched resonantly interacting Bose gas}
%\title{Fate of Hanbury Brown and Twiss correlations in a resonantly interacting Bose gas}

\author{V. E. Colussi}
\email[Corresponding author:  ]{colussiv@gmail.com}
\affiliation{Eindhoven University of Technology, PO Box 513, 5600 MB Eindhoven, The Netherlands}
\author{B. E. van Zwol}
\affiliation{Eindhoven University of Technology, PO Box 513, 5600 MB Eindhoven, The Netherlands}
\author{J. P. D'Incao}
\affiliation{JILA, NIST and Department of Physics, University of Colorado, Boulder, Colorado 80309-0440, USA}
\author{S. J. J. M. F. Kokkelmans}
\affiliation{Eindhoven University of Technology, PO Box 513, 5600 MB Eindhoven, The Netherlands}

\begin{abstract}  
We study the growth of two- and three-body correlations in an ultracold Bose gas quenched to unitarity.  This is encoded in the dynamics of the two- and three-body contacts analyzed in this work.  Via a set of relations connecting many-body correlations dynamics with few-body models, signatures of the Efimov effect are mapped out as a function of evolution time at unitarity over a range of atomic densities $n$.  For the thermal resonantly interacting Bose gas, we find that  atom-bunching leads to an enhanced growth of few-body correlations.  These atom-bunching effects also highlight the interplay between few-body correlations that occurs before genuine many-body effects enter on Fermi timescales.    
\end{abstract}

\maketitle 
\section{Introduction}\label{sec:intro}
Properties of the ultracold Bose gas in the unitary regime ($n|a|^3\to\infty$) are paradigmatic for other strongly-correlated systems where the $s$-wave scattering length, $a$, is much larger than the range of interactions \cite{0034-4885-72-12-126001,PhysRevLett.91.102002}.  However, experimental studies of this regime are severely limited by the enhanced decay of pairs of atoms into deeply bound states leading to loss rates scaling as $\dot{n}/n\propto n^2a^4$. Recently, experimental control of $a$ near a Feshbach resonance \cite{chin2017ultracold} has opened up a pathway to creating and exploring the unitary Bose gas by rapidly quenching interactions to unitarity $|a|\to\infty$.  The inherent metastability of this system due to the Efimov effect \cite{efimov1979low,BRAATEN2006259,WANG20131,d2017few,klauss2017observation} combined with the emergence of a prethermal state
 \cite{PhysRevLett.93.142002,PhysRevA.93.033653,eigen2017universal,eigen2018prethermal} have presented a theoretical puzzle as the relevance of ground-state predictions to this nonequilibrium gas remains unclear \cite{PhysRevLett.108.195301,piatecki2014efimov,PhysRevA.89.041602,PhysRevLett.119.223002,PhysRevLett.103.025302,PhysRevA.95.053602,PhysRevA.97.033621,sze2018two}. 

 The situation is radically different for the unitary two-component Fermi gas, which is comparatively stable due to Pauli suppression of losses \cite{PhysRevLett.93.090404}.  Studies of this system over the last two decades \cite{zwerger2011bcs} have confirmed a universal thermodynamics parametrized by the ``Fermi'' scales $k_\mathrm{n}=(6\pi^2 n)^{1/3}$, $E_\mathrm{n}=\hbar^2k_\mathrm{n}^2/2m$, $t_\mathrm{n}=\hbar/E_\mathrm{n}$ \cite{PhysRevLett.92.090402} where $m$ is the atomic mass.  In addition, a parameter referred to as the two-body contact is central to a set of universal relations between system properties as derived by Tan \cite{TAN20082952,TAN20082971,TAN20082987}.  For bosonic systems, the Efimov effect introduces additional relations involving the three-body contact parameter\cite{PhysRevA.86.053633,PhysRevLett.106.153005}.  The two- and three-body contacts determine the behavior of pair and triplet correlations as two and three atoms approach one another, respectively.  The addition of the three-body contact underscores the underlying intrinsic discrete-scaling of the unitary Bose gas \cite{chin2017ultracold} and the complex role of non-perturbative few-body physics in dictating properties of the gas.   

Over the past few years, experiments have begun to explore Bose gases at unitarity by rapidly quenching from weak interactions to the unitary regime effectively circumventing fast atomic losses over a limited window of time \cite{makotyn2014universal,klauss2017observation,eigen2017universal,eigen2018prethermal,Fletcher377}.  By studying the post-quench dynamics of the contacts, the interplay of few-body correlations in the unitary Bose gas may be unraveled.  This was done interferometrically by Fletcher {\it et al.}~\cite{Fletcher377} in the nondegenerate regime by measuring Ramsey oscillations and extracting the two- and three-body contacts in different regions of the cloud.  In this work, we denote such locally defined contacts as $\mathcal{C}_2$ and $\mathcal{C}_3$.  The integration of the intensive $\mathcal{C}_2$ and $\mathcal{C}_3$ over the extent of the system yields the extensive contacts $C_2$ and $C_3$, respectively \cite{PhysRevA.86.053633,PhysRevLett.106.153005}.  In the unitary regime, it was observed that the two-body contact saturated too quickly to be time-resolved and that the three-body contact slowly approached an equilibrium prediction \cite{PhysRevLett.110.163202,PhysRevLett.112.110402}.  In a degenerate Bose gas, contact dynamics have not been measured directly, however, the tail of the single-particle momentum distribution observed by Makotyn {\it et al.}~\cite{makotyn2014universal} has been argued as consistent with nonzero contacts \cite{PhysRevLett.112.110402,PhysRevA.92.062716} although they were not observed in the time-resolved measurements of Ref.~\cite{eigen2018prethermal}.

Theoretically modeling the experimental quench sequence and subsequent dynamics at unitarity is non-trivial in this strongly-correlated regime.  However, the existence of exact solutions of the unitary few-body problem provides an alternative pathway to studying the many-body dynamics by extracting the dynamical contacts.  This is made possible through a set of short-distance relations derived in Refs.~\cite{PhysRevA.86.053633,PhysRevA.91.013616,PhysRevLett.120.100401} that collectively relate the dynamics of few-body wave functions with the correlation functions and contacts.  In Ref.~\cite{PhysRevA.91.013616}, these relations, in conjunction with solutions of the unitary two-body problem, yielded universal leading-order dynamics of $\mathcal{C}_2$ in a degenerate Bose gas that agree {\it quantitatively} with a many-body model including up to two-body correlations \cite{PhysRevA.89.021601}.  The insensitivity of this agreement to the long-range details of the few-body model employed (trapped, untrapped, etc...) was demonstrated in Ref.~\cite{PhysRevA.91.013616}.  Physically, this is due to the isolation of early-time correlation growth $(t<t_\mathrm{n})$ from long-range physics on scales comparable and larger than $n^{-1/3}$.  These relations were also used to make a robust prediction for leading-order dynamics of $\mathcal{C}_3$ in the degenerate regime \cite{PhysRevLett.120.100401}.  These dynamics depend log-periodically on both $n$ and the three-body parameter $\kappa_*$, which is the wave-number of the ground-state Efimov trimer \cite{efimov1979low,BRAATEN2006259}. 
 
 In principle these short-distance relations hold also for the nondegenerate Bose gas quenched to unitarity with some additional qualifications.  In this regime, it is not immediately clear how to parametrize the nonequilibrium dynamics in terms of the Fermi and thermal scales that includes the thermal de Broglie wavelength $\lambda=(2\pi\hbar^2/mk_b T)^{1/2}$ and time $t_\lambda=\hbar/k_bT$.  However,  recent experimental results in this regime have highlighted the central role played by the geometric mean referred to as the ``meeting-time'' $t_s=(t_\mathrm{n}t_\lambda)^{1/2}$, which measures the travel time for an atom moving at the thermal speed to reach its neighbors \cite{eigen2018prethermal}.  Such an event cannot be properly captured within a few-body model.   However, for Bose gases prepared near the critical point for condensation $n\lambda^3\sim 1$, the timescale matching $t_s\sim t_\mathrm{n}\sim t_\lambda$ allows the contact dynamics to be predicted via few-body models in the same basic spirit as Refs.~\cite{PhysRevA.91.013616,PhysRevLett.120.100401}.  Experimentally, this phase-space density range was achieved near trap center in Ref.~\cite{Fletcher377} and spans part of the range explored in Ref.~\cite{eigen2018prethermal}.  At lower phase-space densities, such predictions can still be made but only at shorter times $t<t_\lambda,t_\mathrm{n}$.  Additionally, we expect that atom-bunching effects due to the Hanbury-Brown-Twiss effect \cite{brown1956correlation,doi:10.1119/1.1937827} may play a role in the contact dynamics, making this an intriguing regime to study.

 In this work, we quench both a pure Bose-Einsein condensate (BEC), approximated as a coherent state, and an ideal thermal Bose gas to unitarity and investigate the subsequent growth of few-body correlations in a uniform system.   These states are chosen to approximate the regimes explored experimentally using $^{85}$Rb and $^{39}$K in Refs.~\cite{makotyn2014universal,klauss2017observation} and \cite{eigen2017universal,eigen2018prethermal}, respectively.  Via analytic solutions of the unitary three-body problem \cite{PhysRevLett.97.150401} and the set of short-distance relations, we extend the study of the dynamics of $\mathcal{C}_3$ in Ref.~\cite{PhysRevLett.120.100401} to the thermal regime where log-periodic signatures are enhanced due to atom-bunching.  We also map out the dynamics of $\mathcal{C}_2$ for both thermal and BEC initial conditions.  $\mathcal{C}_2$ measures the number of pairs per (volume)$^{4/3}$ \cite{PhysRevA.78.053606}, however we find that after an initial period of universal evolution the number of pairs becomes sensitive to the surrounding ``medium'' consisting of the third boson.    Crucially, through this medium effect, $\mathcal{C}_2$ develops a {\it secondary} dependence on the Efimov effect prior to genuine many-body effects which enter on Fermi timescales.  Due to the lack of a fourth boson in our model, there is no analogous medium effect in the dynamics of $\mathcal{C}_3$, which measures the number of triplets per (volume)$^{5/3}$  \cite{PhysRevA.78.053606}.  The dependence of $\mathcal{C}_3$ on the Efimov effect in our model and in Ref.~\cite{PhysRevLett.120.100401} is therefore {\it primary}.  Additionally, by comparing the post-quench dynamics for both thermal and BEC initial states and searching for multiplicative signatures due to atom-bunching, we further highlight the emergence of this medium effect in the dynamics of $\mathcal{C}_2$.    

This paper is organized as follows.  We begin in Sec.~\ref{sec:connections} by reviewing connections between the many-body correlation dynamics and few-body models in a quenched uniform gas.  These relations are then appropriately calibrated to model a Bose gas whose initial state is either BEC or thermal.  Via these connections, we employ analytic solutions to the unitary three-body problem to study the post-quench dynamics of $\mathcal{C}_2$ in Sec.~\ref{sec:c2} and $\mathcal{C}_3$ in Sec.~\ref{sec:c3}.  We conclude in Sec.~\ref{sec:conclusion}, commenting on experimental implications throughout.  In the Appendix we provide additional details related to the calculation for completeness.  In Sec.~\ref{sec:appendix}, analytic solutions for three resonantly interacting bosons in a trap given in Ref.~\cite{PhysRevLett.97.150401} are outlined, and we provide technical derivations of results used in this work specific to these solutions.  In Sec.~\ref{sec:convergence} we provide details related to the convergence of our calculations.    

\section{Many-body Correlation dynamics via few-body models.}\label{sec:connections}

In this section, we outline a set of short-distance relations between few-body wave functions, correlation functions, and contacts at unitarity, first given in Refs.~\cite{PhysRevA.91.013616,PhysRevLett.120.100401,PhysRevLett.110.163202}.  For alkali atoms, the species dependent van der Waals length, $r_\mathrm{vdW}$, furnishes a natural short-range for interparticle interactions \cite{RevModPhys.82.1225}.  In an ultracold Bose gas, the typical momentum scale is such that $kr_\mathrm{vdW}\ll 1$ and the $s$-wave scattering length determines the low-energy physics, captured by the zero-range model.  The connections outlined in this section are made at distances larger than $r_\mathrm{vdW}$ but shorter than the scales $a$, $n^{-1/3}$, and $\lambda$.

We begin by relating the short-distance behavior of few-body correlation functions to the contacts.  The normalized pair correlation function is defined in second-quantization as \cite{cohen2011advances}
\begin{equation}\label{eq:g2def}
g^{(2)}({\bf r},{\bf r'})\equiv\frac{\langle \hat{\psi}^\dagger({\bf r})  \hat{\psi}^\dagger({\bf r'})  \hat{\psi}({\bf r'})  \hat{\psi}({\bf r})\rangle}{\langle\hat{\psi}^\dagger({\bf r})\hat{\psi}({\bf r})\rangle \langle\hat{\psi}^\dagger({\bf r'})\hat{\psi}({\bf r'})\rangle},
\end{equation}
in terms of the bosonic field operators $\hat{\psi}({\bf r})=\sum_{\bf k}\hat{a}_{\bf k}e^{i{\bf k}\cdot{\bf r}}/\sqrt{V}$ and system volume $V$.  In a uniform system the first-order correlation function $\langle\hat{\psi}^\dagger({\bf r})\hat{\psi}({\bf r})\rangle=N/V=n$.  Assuming translational invariance, we ignore the center of mass dependence of a pair and write $g^{(2)}({\bf r})\equiv g^{(2)}({\bf 0},{\bf r})$.  In first quantization, the numerator of Eq.~\eqref{eq:g2def} is equivalent to the trace over all other coordinates $N(N-1)|\langle \Psi_\mathrm{MB}|(|{\bf 0}\rangle_1\otimes|{\bf r}\rangle_2)|^2$ for indistinguishable bosons \cite{PhysRevA.86.053633}.  This gives
\begin{align}\label{eq:g2reldef}
g^{(2)}({\bf r})&=\frac{N(N-1)}{n^2}\int d^3 r_3\dots d^3 r_N\nonumber\\
&\quad\quad\times |\Psi_\mathrm{MB}({\bf 0},{\bf r},{\bf r_3},\dots,{\bf r}_N)|^2.
\end{align}
As $r$ vanishes, the many-body wave function behaves as $\Psi_\mathrm{MB}({\bf 0},{\bf r},{\bf r_3},\dots {\bf r}_N)\approx \mathcal{A}({\bf c},\{{\bf r}_l\}_{l>2})/r$, where $1/r$ matches the functional behavior of the zero-energy two-body scattering state at unitarity in the zero-range model \cite{PhysRevA.86.053633}.  The function $\mathcal{A}$ depends on the distinct positions of the remaining atoms and the center of mass of the interacting pair ${\bf c}={\bf r}/2$.  In this limit, Eq.~\eqref{eq:g2reldef} becomes \cite{PhysRevA.86.053633}
\begin{align}
g^{(2)}({\bf r})&\underset{{r}\rightarrow 0}{=}\frac{N(N-1)}{n^2r^2}\int d^3 r_3\dots d^3 r_N |\mathcal{A}({\bf c},\{{\bf r}_l\}_{l>2})|^2,\label{eq:g2reldeflim}\\
&\underset{{r}\rightarrow 0}{=}\frac{\mathcal{C}_2}{16\pi^2n^2r^2},\label{eq:c2g2}
%&\underset{{r}\rightarrow 0}{=}\frac{1}{n^2r^2}\langle \mathcal{A},\mathcal{A}\rangle({\bf c}).
\end{align}
where we have dropped the dependence on ${\bf c}$ due to translational invariance.  

We now outline a relation analogous to Eq.~\eqref{eq:c2g2} for the behavior of the normalized triplet correlation function as the separation between three bosons vanishes.  In second-quantization, the normalized triplet correlation function is defined as \cite{cohen2011advances}
\begin{equation}\label{eq:g3def}
g^{(3)}({\bf r},{\bf r'},{\bf r'''})\equiv\frac{\langle \hat{\psi}^\dagger({\bf r})  \hat{\psi}^\dagger({\bf r'})  \hat{\psi}^\dagger({\bf r''})   \hat{\psi}({\bf r''}) \hat{\psi}({\bf r'})  \hat{\psi}({\bf r})\rangle}{\langle\hat{\psi}^\dagger({\bf r})\hat{\psi}({\bf r})\rangle \langle\hat{\psi}^\dagger({\bf r'})\hat{\psi}({\bf r'})\rangle \langle\hat{\psi}^\dagger({\bf r''})\hat{\psi}({\bf r''})\rangle}.
\end{equation}
 In first quantization, the numerator of Eq.~\eqref{eq:g3def} is equivalent to the trace over all other coordinates $N(N-1)(N-2)\langle \Psi_\mathrm{MB}|(|{\bf 0}\rangle_1\otimes|{\bf r}\rangle_2\otimes|{\bf r'}\rangle_3)$ for identical bosons in a translationally invariant system \cite{PhysRevA.86.053633}.  This gives  
\begin{align}\label{eq:g3reldef1}
g^{(3)}({\bf 0},{\bf r},{\bf r'})=&\frac{N(N-1)(N-2)}{n^3}\int d^3 r_4\dots d^3 r_N\nonumber\\
&\times |\Psi_\mathrm{MB}({\bf 0},{\bf r},{\bf r'},{\bf r}_4\dots,{\bf r}_N)|^2.
\end{align}
In a uniform system, translational invariance allows us to ignore center of mass dependence and write $g^{(3)}(R,{\bf \Omega})\equiv g^{(3)}({\bf 0},{\bf r},{\bf r'})$ in terms of the hyperradius $R\equiv\sqrt{r^2+\rho^2}/2$ written in terms of the Jacobi vectors ${\bf r}$ and $\boldsymbol{\rho}\equiv(2{\bf r}'-{\bf r})/\sqrt{3}$ and the hyperangles ${\bf \Omega}=\{\hat{\rho},\hat{r},\alpha\}$.  The set of hyperangles consists of the spherical angles for each Jacobi vector and the hyperangle $\tan(\alpha)=r/\rho$ \cite{convention}.  In the limit $R\to 0$, where the separation between three-bosons vanishes with fixed hyperangles, the many-body wave function behaves as $\Psi_\mathrm{MB}({\bf 0},{\bf r},{\bf r'},{\bf r}_4,\dots {\bf r}_N)\approx \mathcal{B}({\bf C},\{{\bf r}_l\}_{l>3})\Psi_{sc}(R,{\bf \Omega})$ where ${\bf C}=({\bf r}+{\bf  r'})/3$ \cite{PhysRevA.86.053633}.  The three-body wave function $\Psi_{sc}(R,{\bf \Omega})$ is the zero-energy three-body scattering state
\begin{equation}
\Psi_{sc}(R,{\bf \Omega})=\frac{1}{R^2}\sin\left[s_0\ln\frac{R}{R_t}\right]\frac{\phi_{s_0}({\bf \Omega})}{\sqrt{\langle\phi_{s_0}|\phi_{s_0}\rangle}},
\end{equation}
depending on Efimov's constant $s_0\approx 1.00624$ \cite{efimov1971weakly}, the three-body parameter $ R_t=\sqrt{2}\exp(\text{Im}\ln[\Gamma(1+is_0)]/s_0)/\kappa_*$ that sets the phase of the log-periodic oscillation, and the hyperangular wave function $\phi_{s_0}({\bf \Omega})=(1+\hat{Q})\sinh(s_0(\pi/2-\alpha)/\sqrt{4\pi}\sin2\alpha$ for three identical bosons in the lowest state of angular momentum.  The operator $\hat{Q}=\hat{P}_{13}+\hat{P}_{23}$ is written in terms of the $\hat{P}_{ij}$'s that permute particles $i$ and $j$.  We refer the reader to App.~\ref{sec:appendix} for analytic expressions of the hyperangular normalization factor $\langle\phi_{s_0}|\phi_{s_0}\rangle$.  In this limit, Eq.~\eqref{eq:g3reldef1} becomes \cite{PhysRevA.86.053633}
\begin{align}
g^{(3)}(R,{\bf \Omega})\underset{{R}\rightarrow 0}{=}&\frac{N(N-1)(N-2)}{n^3}|\Psi_{sc}(R,{\bf\Omega})|^2\nonumber\\
&\times\int d^3 r_4\dots d^3 r_N |\mathcal{B}({\bf C},\{{\bf r}_l\}_{l>3})|^2,\label{eq:g3reldef}\\
g^{(3)}(R,{\bf \Omega})\underset{{R}\rightarrow 0}{=}&|\Psi_{sc}(R,{\bf\Omega})|^2\frac{8}{n^3 s_0^2\sqrt{3}}\mathcal{C}_3,\label{eq:c3g3}
%\underset{{R}\rightarrow 0}{=}&|\Psi_{sc}(R,{\bf\Omega})|^2\frac{\langle \mathcal{B},\mathcal{B}\rangle({\bf C})}{n^3}.
\end{align}
where we have dropped the dependence on ${\bf C}$ due to translational invariance.  

We now relate the contacts to the short-distance behavior of three-body wave functions after a quench.  To derive these relations, we review the interpretation of correlation functions as conditional probabilities \cite{pathria}.  If we measure an atom whose position defines the origin, $ng^{(2)}({\bf r})$ is the conditional probability density for measuring another atom in a volume $dV$ ($ndV\ll1$) at ${\bf r}$ \cite{pathria}.  The magnitude of $\mathcal{C}_2$ therefore dictates the probability for measuring correlated pairs.  Analogously, $n^2 g^{(3)}(R,{\bf \Omega})$ is the conditional probability for finding two other atoms in volume elements $dV$ whose locations are defined by the three-body configuration ($R,{\bf \Omega}$) \cite{pathria}.  The magnitude of $\mathcal{C}_3$ therefore dictates the probability for measuring correlated triplets.  In a three-body model, these probability densities can be extracted from a calibrated three-body wave function $\Psi(R,{\bf \Omega},t)$ via the following relations \cite{PhysRevLett.120.100401}
\begin{align}
ng^{(2)}({\bf r},t)&\underset{{|{\bf r}|}\rightarrow 0}{=}2\int d^3 r_{3,12}|\Psi({\bf r},\boldsymbol{\rho},t)|^2,\label{eq:g2psi2}\\
n^2 g^{(3)}(R,{\bf \Omega},t)&\underset{{R}\rightarrow 0}{=}2|\Psi(R,{\bf \Omega},t)|^2,\label{eq:g3psi3}
\end{align}    
where ${\bf r}_{3,12}=\boldsymbol{\rho}\sqrt{3}/2$, and the factor of 2 in Eqs.~\eqref{eq:g2psi2} and \eqref{eq:g3psi3} arises due to indistinguishability of the measured atoms.  Equations \eqref{eq:c2g2}, \eqref{eq:c3g3}--\eqref{eq:g3psi3} constitute the basic short-distance relations referenced in Sec.~\ref{sec:intro} serving as the foundation of our analysis of the many-body correlation dynamics in this work.  

Before moving on, we comment on the sense in which Eqs.~\eqref{eq:g2psi2} and \eqref{eq:g3psi3} may be used to make predictions for the two- and three-body correlations in a quenched ultracold Bose gas.  The quench immediately impacts the behavior of the many-body wave function at short distances, which becomes singular as $1/r$.  Subsequently, there is a lag between the early-time correlation growth ($t\ll t_\mathrm{n}$) at short-distance resulting from this disturbance, and the evolution of the bulk medium on  timescales $t_\mathrm{n}$, $t_s$, and $t_\lambda$.  In this picture, the character of the early-time growth of correlations at short distances is therefore few-body in nature.  Not all observables however may show a clear signature of the contacts as evidenced in the recent experimental observation of an exponential rather than the expected \cite{TAN20082952} power law form of high-momentum tail of the single-particle momentum distribution \cite{eigen2018prethermal}.  Additionally, the observed $k$-dependent rate at which the occupation of high-momentum modes plateau in that work suggests a competition between short-distance few-body physics and the equilibrating effect of quasiparticle collisions in this particular observable \cite{PhysRevA.98.053612}.  Quantitatively understanding relaxation and equilibration dynamics for quenched ultracold gases remains an important, difficult, and open topic that we leave for future study \cite{PhysRevA.90.021602,Hung1237557}. 
%should I make a note about xiao's work?
%
%
%
%emphasize that this calculation is NOT sensitive to the trap at early times.  This has been confused by Doerte.  
\subsection{Initial Conditions}
In order to predict contact dynamics from a three-body wave function, it is necessary to choose initial conditions such that Eqs.~\eqref{eq:g2psi2} and \eqref{eq:g3psi3} matches the prepared many-body system under study.  In this work, we consider an ultracold Bose gas prepared as either a BEC or an ideal thermal Bose gas.  The limiting behavior of correlation functions in these cases is \cite{cohen2011advances}
\begin{equation}\label{eq:gzero}
g^{(l)}({\bf r},\dots,{\bf r})=\begin{cases}
														1\quad\quad(\text{BEC})\\
														l!\quad\quad(\text{Thermal})\\
											     \end{cases}.
\end{equation}
Physically, the $l!$ bunching factor means that if we measure one boson at the location ${\bf r}$ the probability of simultaneously finding $l-1$ additional identical bosons at this same location is $l!$ more likely than the individual probabilities for measuring each boson at this place.  For a BEC approximated as a coherent state, this probability is always the same regardless of the number of bosons \cite{PhysRev.131.2766}.    

In this work, we follow Ref.~\cite{PhysRevLett.120.100401} and use the initial three-body wave function 
\begin{equation}\label{eq:psi0}
\Psi(R,{\bf \Omega},0)=Ae^{-R^2/2B_1^2}\left[1-\left(\frac{R}{B_2}\right)^2\right],
\end{equation}
where $A$ is the normalization constant 
\begin{equation}
A^{-2}=(3\pi^2)^{3/2}\left(B_1^6+12\frac{B_1^{10}}{B_2^4}-6\frac{B_1^8}{B_2^2}\right),
\end{equation}
so that $\langle \Psi|\Psi\rangle=1$.  The lengths $B_1$ and $B_2$ must be carefully chosen so that the initial boundary conditions
\begin{align}
2\int d^3 r_{3,12}|\Psi({\bf r},\boldsymbol{\rho},0)|^2/n&\underset{{|{\bf r}|}\rightarrow 0}{=}\begin{cases}
														1\quad\quad(\text{BEC})\\
														2!\quad\quad(\text{Thermal})\\
											     \end{cases},\label{eq:g2psi2init}\\
2|\Psi(R,{\bf \Omega},0)|^2/n^2&\underset{{R}\rightarrow 0}{=}\begin{cases}
														1\quad\quad(\text{BEC})\\
														3!\quad\quad(\text{Thermal})\\
											     \end{cases},\label{eq:g3psi3init}
\end{align}
%\begin{align}
%ng^{(2)}({\bf r},0)&\underset{{|{\bf r}|}\rightarrow 0}{=}2\int d^3 r_{3,12}|\Psi({\bf r},\boldsymbol{\rho},0)|^2,\label{eq:g2psi2init}\\
%n^2 g^{(3)}(R,{\bf \Omega},0)&\underset{{R}\rightarrow 0}{=}2|\Psi(R,{\bf \Omega},0)|^2,\label{eq:g3psi3init}
%\end{align}    
are satisfied.  We find that the choices $k_\mathrm{n}B_1\approx2.3422,\ k_\mathrm{n}B_2\approx 4.3959$ and $k_\mathrm{n}B_1\approx 1.6679,\ k_\mathrm{n}B_2\approx 2.8816$ match the initial boundary conditions for a BEC and an ideal thermal Bose gas, respectively.  

It is tempting to interpret the long-range details of Eq.~\eqref{eq:psi0} and lengths $B_1$ and $B_2$ physically, e.g. in terms of an artificial trap.  However, it was shown in Refs.~\cite{PhysRevA.91.013616,PhysRevLett.120.100401} that provided the initial boundary conditions [Eqs.~\eqref{eq:g2psi2init} and \eqref{eq:g3psi3init}] are satisfied, the obtained early-time contact dynamics are largely insensitive to long-range details like the presence or absence of an artificial trap.  In these works, this insensitivity was shown to be robust to both variation of the functional form of the initial wave few-body wave function and variation of the frequency of an artificial trap.  

\subsection{Unitary Three-Body Problem}\label{sec:res3b}

For convenience, we have chosen to use analytic solutions of the trapped unitary three-body problem in the zero-range model given in Ref.~\cite{PhysRevLett.97.150401}, although any set of three-body eigenstates (free-space, trapped, box, etc...) at unitarity may be used to study the contact dynamics.  Due to the arbitrariness of this choice, we focus here only on general features of the unitary three-body problem, moving specifics of our calculations related to the chosen three-body basis to the Appendices in Apps.~\ref{sec:appendix} and \ref{sec:convergence}.  

We follow the general approach of Efimov \cite{efimov1971weakly} and factor unitary three-body relative eigenstates as $|\Psi_{s,j}\rangle=\mathcal{N}_{s,j}F_j^{(s)}(R)\phi_s({\bf \Omega})/R^2$ where $\mathcal{N}_{s,j}^{-2}=\langle \Psi_{s,j}|\Psi_{s,j}\rangle$ is the normalization factor.  The hyperrangular eigenstates $\phi_s({\bf \Omega})=(1+\hat{Q})\varphi_s(\alpha)/\sqrt{4\pi}\sin2\alpha$ solve the hyperangular eigenvalue problem \cite{efimov1979low}
\begin{align}
\varphi_s''(\alpha)&=s^2\varphi_s(\alpha),\\
\varphi_s'(0)&=\frac{8}{\sqrt{3}}\varphi_s(\pi/3)=0,\\
\varphi_s(\pi/2)&=0,
\end{align}
yielding $\varphi_s(\alpha)=\sin(s(\pi/2-\alpha)$ and channel labels $s=is_0,s_1\dots$ that are solutions of the transcendental equation
\begin{equation}
\frac{8}{\sqrt{3}}\sin\left(\frac{s\pi}{6}\right)-s\cos\left(\frac{s\pi}{2}\right)=0.
\end{equation}
 For channels with $s^2>0$, the $R\to0$ behavior of the hyperradial eigenstates is $F_j^{(s)}(R)\propto O(R^s)$.  In the Efimov channel where $s=is_0$, an additional boundary condition $F_j^{(is_0)}(R)\propto \sin[s_0\ln(R/R_t)]$ is required to preserve the Hermiticity of the problem \cite{danilov1961gs,PhysRevLett.97.150401}.  
 
 This model can be extended to include three-body losses by simply modifying the three-body parameter dependence in this boundary condition as $\ln R_t\to\ln R_t-i\eta/s_0$ to imitate flux loss to deeply-bound molecular decay channels \cite{PhysRevA.86.053633,PhysRevA.67.022505}.  For $^{85}$Rb \cite{PhysRevLett.108.145305} and $^{39}$K \cite{PhysRevLett.111.125303}, the experimentally measured inelasticity parameters are $\eta=0.06$ and $\eta=0.09$, respectively.  By taking a derivative of the eigenenergies $E_\mathrm{3b}^{(j)}$ in the Efimov channel, the widths $\Gamma_j$ can be obtained to first order in $\eta$ \cite{PhysRevA.86.053633}
%notice hte extra factor of \hbar different from our prl, where we got it wrong.  difference between the decay rate and the width. 
\begin{equation}\label{eq:widths}
\Gamma_j\underset{{\eta}\rightarrow 0}{=}\eta \frac{4\hbar^2}{ms_0}C_3^{(j)},
\end{equation}  
where the extensive three-body contact for each trapped eigenstate are calculated analytically in App.~\ref{sec:appendix}.  The time-dependent unitary three-body wave function is obtained by projection onto the initial wave function, Eq.~\eqref{eq:psi0}, via the sudden approximation
\begin{equation}\label{eq:3bpsi}
\Psi(R,\Omega,t)=\sum_{s,j}c_{s,j}\Psi_{s,j}(R,{\bf \Omega})e^{-iE_\mathrm{3b}^{(j)}t/\hbar}e^{-\Gamma_j t/2\hbar}    
\end{equation}
where the summation runs over all channels and quantum numbers and includes the overlaps $c_{s,j}\equiv\langle \Psi_{s,j}|\Psi_0\rangle$ [see App.~\ref{sec:overlaps} for analytic expressions.]  We further classify the physics described by each channel as universal ($s>0$) or nonuniversal ($s=is_0$) due to independence or dependence, respectively, on the three-body parameter $\kappa_*$.

\section{Post-quench dynamics of $\mathcal{C}_2(t)$.}\label{sec:c2}

In this section, we study the early-time dynamics of $\mathcal{C}_2(t)$ for BEC and thermal initial conditions, which amounts to enforcing Eq.~\eqref{eq:gzero}.  This is accomplished by substituting our three-body wave function, Eq.~\eqref{eq:3bpsi}, into Eq.~\eqref{eq:g2psi2} to link the growth of pair correlations in the gas with the short-distance behavior of the three-body model outlined in Sec.~\ref{sec:res3b}.  Crucially, the presence of a third boson external to a pair plays a role here analogous to the medium, introducing secondary Efimov and bunching effects into the dynamics of $\mathcal{C}_2$ as its presence is increasingly felt at later times.  This results in dynamics of $\mathcal{C}_2(t)$ that depart from universal predictions found in the literature \cite{PhysRevA.89.021601,PhysRevA.91.013616} before many-body effects take place.  

We begin by deriving an expression for $\mathcal{C}_2(t)$ using the three-body wave function in Eq.~\eqref{eq:3bpsi}.  To emphasize the generality of this approach, results specific to the particular basis of three-body eigenstates used are located in the App.~\ref{sec:appendix} for reasons of clarity and completeness.   We begin by rewriting the three-body eigenfunctions as $\Psi_{s,j}(R,{\bf \Omega})=(1+\hat{Q})\chi_{s,j}({\bf r},\boldsymbol{\rho})/r\rho$, where the $\chi$ functions are finite in the limit ${\bf r}\to0$ and vanish in the limit $|\boldsymbol{\rho}|\to 0$ \cite{0034-4885-80-5-056001}.  Combining Eqs.~ \eqref{eq:c2g2} and \eqref{eq:g2psi2} gives the following
\begin{equation}\label{eq:c2step1}    
\frac{\mathcal{C}_2(t)}{32\pi^2n}\underset{{|{\bf r}|}\rightarrow 0}{=}\int d^3 r_{3,12} \left|r\sum_{s,j}c_{s,j}(t)(1+\hat{Q})\frac{\chi({\bf r},\boldsymbol{\rho})}{r\rho}\right|^2,
\end{equation}
where we have included the time-dependent phase factors in the overlaps $c_{s,j}(t)\equiv c_{s,j}\exp[-i(E_\mathrm{3b}^{(j)}-i\Gamma_j/2)t/\hbar]$ for notational simplicity. In the limit $|{\bf r}|\to 0$, the Jacobi vectors can be related via kinematic relations  \cite{0034-4885-80-5-056001} to give
\begin{equation}
\hat{Q}\left[\frac{\chi({\bf r},\boldsymbol{\rho})}{r\rho}\right]\underset{{|{\bf r}|}\rightarrow 0}{=}\frac{\chi\left(-\frac{\sqrt{3}\boldsymbol{\rho}}{2},-\frac{\boldsymbol{\rho}}{2}\right)}{\sqrt{3}\rho^2/4}-\frac{\chi\left(\frac{\sqrt{3}\boldsymbol{\rho}}{2},-\frac{\boldsymbol{\rho}}{2}\right)}{\sqrt{3}\rho^2/4}.
\end{equation}  
The limit in Eq.~\eqref{eq:c2step1} can now be taken with result
\begin{equation}    
\frac{\mathcal{C}_2(t)}{32\pi^2n}=\int d^3 r_{3,12} \left|\sum_{s,j}c_{s,j}(t)\frac{\chi(0,\boldsymbol{\rho})}{\rho}\right|^2.
\end{equation}
We now recast the $|{\bf r}|\to0$ limit as an equivalent $\alpha\to 0$ limit to obtain 
\begin{align}
\mathcal{C}_2(t)=&32\pi^2n\sum_{s,s',j,j'}c_{s,j}(t)c^*_{s',j'}(t)\mathcal{N}_{s,j}\mathcal{N}^*_{s',j'}\phi_s^0\left[\phi_{s'}^0\right]^*\nonumber\\
&\times\int d^3 r_{3,12} \frac{F^{(s)}_j(R)\left[F^{(s')}_{j'}(R)\right]^*}{2R^2},\label{eq:c2step2}
\end{align}
where $R=r_{3,12}\sqrt{2/3}$ when $r=0$, and we have utilized the shorthand notation $\phi_s^0=\sin(s\pi/2)/\sqrt{4\pi}$.  Using the trapped eigenfunctions of Ref.~\cite{PhysRevLett.97.150401} as discussed in Sec.~\ref{sec:res3b}, all integrals in Eq.~\eqref{eq:c2step2} may be evaluated analytically, and these expressions are calculated in App.~\ref{sec:Atens}.  

\subsection{BEC}\label{sec:c2coherent}
\begin{figure}[h!]
\centering
\includegraphics[width=8.6cm]{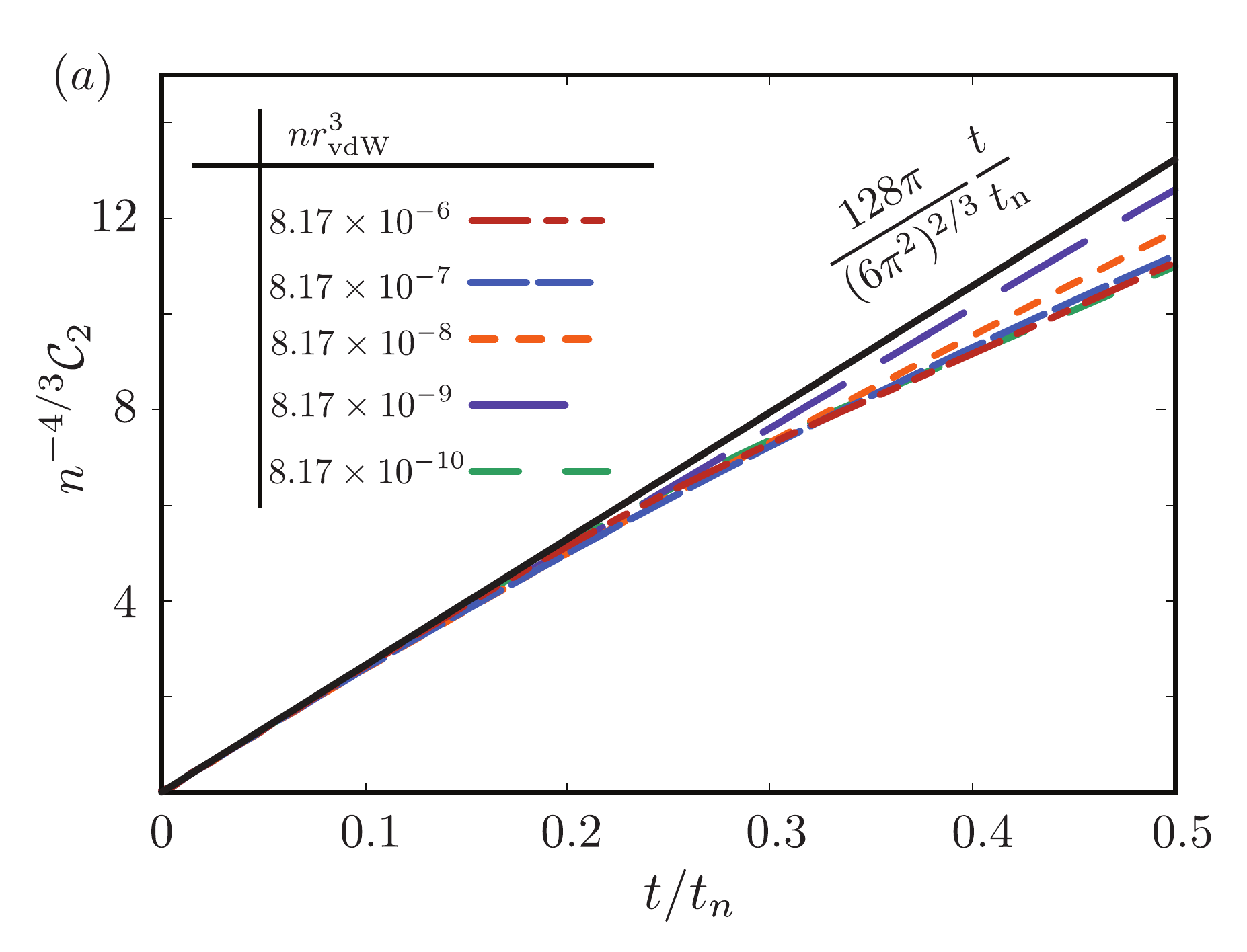}
\includegraphics[width=8.6cm]{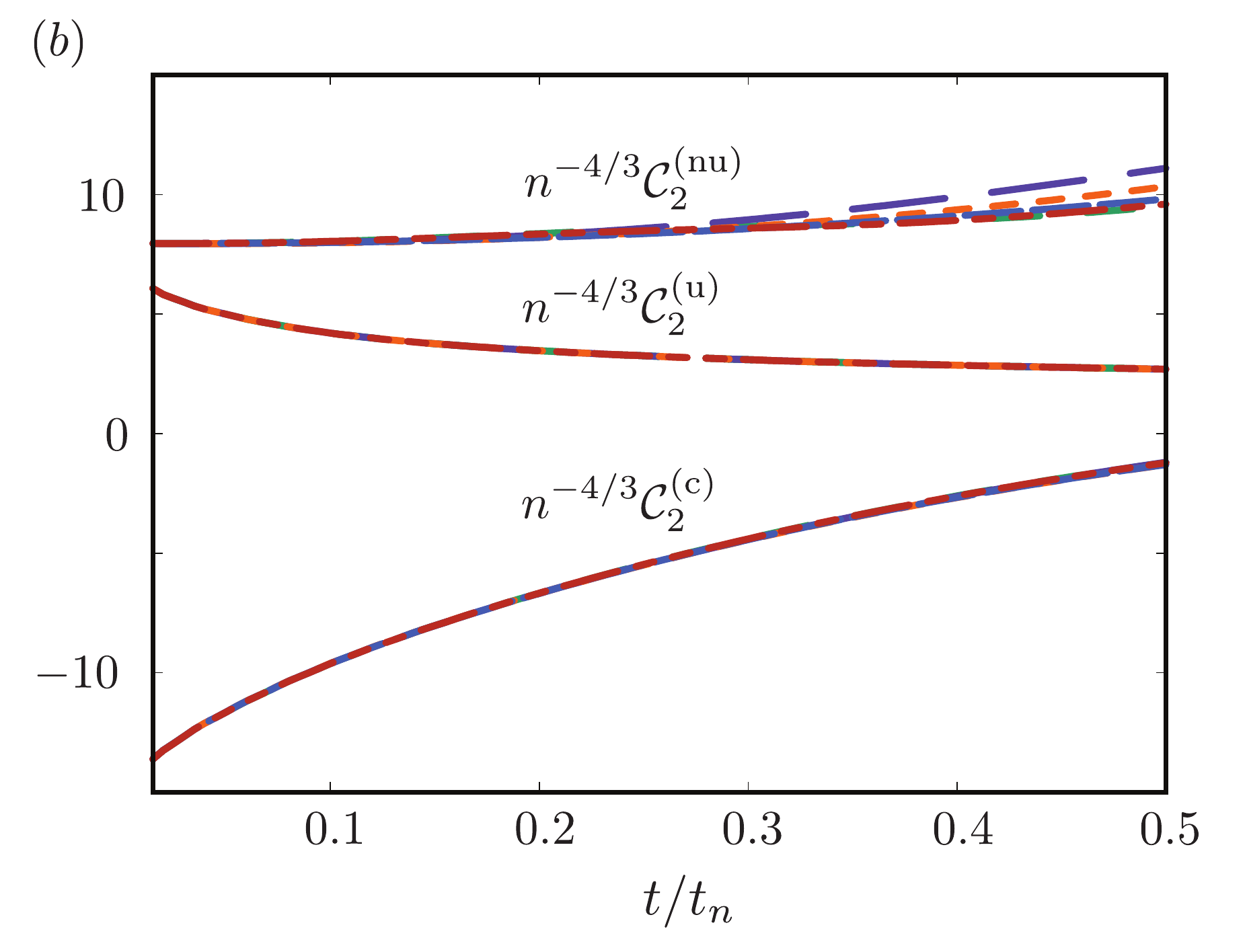}
\includegraphics[width=8.6cm]{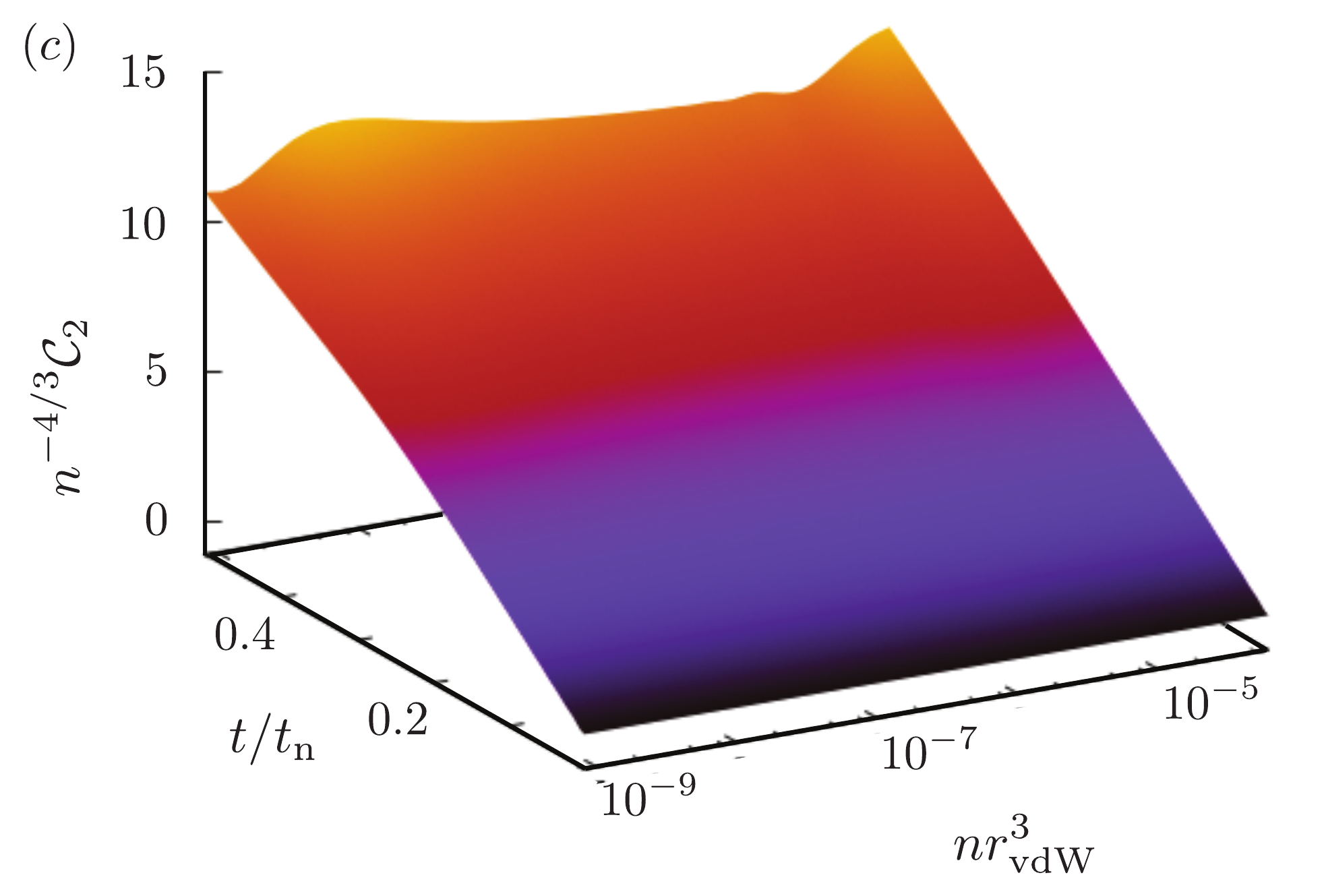}
\caption{(a) Dynamics of $\mathcal{C}_2(t)$ for $\eta=0.06$ for BEC initial conditions over a range of densities of experimental interest compared to the universal leading-order behavior [Eq.~\eqref{eq:c2limitcoherent}.]  (b) Evolution of the components of $\mathcal{C}_2(t)$ for the same parameters as (a).  (c) Dynamical surface showing the evolution of $\mathcal{C}_2(t)$ over a range of densities and times.  
\label{fig:c2coherent}}
\end{figure}
In this section, we present results for $\mathcal{C}_2(t)$ for a BEC quenched to unitarity.  In our model, this amounts to evaluating Eq.~\eqref{eq:c2step2} using BEC initial conditions given in Eq.~\eqref{eq:gzero}.  The leading-order dependence of $\mathcal{C}_2(t)$ in this scenario was derived analytically in Ref.~\cite{PhysRevA.91.013616}
\begin{equation}\label{eq:c2limitcoherent}
n^{-4/3}\mathcal{C}_2(t)=\frac{128\pi}{(6\pi^2)^{2/3}}\frac{t}{t_\mathrm{n}},
\end{equation} 
by solving the unitary two-body problem, agreeing with a many-body variational prediction to within less than 2$\%$.  

Our results for $\mathcal{C}_2(t)$ are shown in Fig.~\ref{fig:c2coherent} over a range of densities of experimental interest.  We find that the initial growth of $\mathcal{C}_2(t)$ follows the universal prediction in Eq.~\eqref{eq:c2limitcoherent} as shown in Fig.~\ref{fig:c2coherent}(a).  Ultimately, this agreement at short times indicates that the presence of a third boson is {\it irrelevant} during this stage.  However, at later times $t\gtrsim0.25$, the dynamics of $\mathcal{C}_2$ include {\it intrinsically} three-body effects:  log-periodic scaling with the atomic density and a beating phenomenon at the frequency of an Efimov trimer.  We focus now on characterizing these effects.     

The log-periodic oscillation of $\mathcal{C}_2(t)$ with atomic density can be seen in the dynamical surface in Fig.~\ref{fig:c2coherent}(c).  This oscillation is relatively small on the order of a $10\%$ variation on top of the continuous scaling of $\mathcal{C}_2(t)$ with the atomic density by $t\sim0.5t_\mathrm{n}$.  By rescaling the atomic density in van der Waals units, the dynamical surface applies to a range of atomic species satisfying $\eta\ll1$, including both $^{85}$Rb and $^{39}$K.  Additionally, the maximum of this surface occurs for densities satisfying 
\begin{equation}\label{eq:resc2coherent}
k_\mathrm{n}R_\mathrm{3b}^{(j)}\approx 0.91,
\end{equation}
where $R_\mathrm{3b}^{(j)}=\sqrt{2(1+s_0^2)/3}\exp(j\pi/s_0)/\kappa_*$ is the size of the $j$th Efimov trimer in free space \cite{BRAATEN2007120}.  This supports the findings of Refs.~\cite{PhysRevLett.120.100401,PhysRevLett.121.023401} that the coincidence of trimer size and interparticle spacing results in correlation enhancement.  

That the dynamics of $\mathcal{C}_2(t)$ should contain a mixture of both universal and nonuniversal characteristics can be seen easily from the coherent sum in Eq.~\eqref{eq:c2step2}, which can be decomposed as $\mathcal{C}_2(t)=\mathcal{C}^{(\mathrm{u})}_2(t)+\mathcal{C}^{(\mathrm{nu})}_2(t)+\mathcal{C}^{(\mathrm{c})}_2(t)$ into contributions which are universal $(s,s'>0)$, nonuniversal $(s=s'=is_0)$, and the remainder which couples universal and Efimov channels, respectively.  Although these components are not physically distinguishable, it is conceptually instructive to analyze their behavior individually as shown in Fig.~\ref{fig:c2coherent}(b).  We see that $n^{-4/3}\mathcal{C}^{(\mathrm{c})}_2(t)$ is to a very good approximation universal, showing virtually no variation with atomic density.  The violation then arises effectively through the dynamics of $n^{-4/3}\mathcal{C}^{(\mathrm{nu})}_2(t)$ for $t\gtrsim0.25t_\mathrm{n}$, which is the only contribution that increases as the system evolves.  This contribution also displays a visible beating phenomenon in time at the frequency of the Efimov trimer with binding energy nearest $E_\mathrm{n}$ such that $|E_\mathrm{3b}^{(j)}|> E_\mathrm{n}$.  This phenomenon was also found in the early-time dynamics of $\mathcal{C}_3(t)$ in Ref.~\cite{PhysRevLett.120.100401}, which we revisit in Sec.~\ref{sec:c3coherent}.  

The delayed appearance of log-periodicities and trimer beating in the pair correlation dynamics is a signature of the secondary influence of the medium played by the third boson.  Importantly, this delay occurs prior to $t\sim t_\mathrm{n}$ where we expect genuine many-body effects to become important.  We return to the picture of the third boson as a medium in Sec.~\ref{sec:c2therm}, where atom-bunching effects allow us to further characterize this metaphor.   
%It would interesting to know whether this variation should show up in the metastable state, so taht the fit of Braaten doesn't work so well?  Probably it gets killed by dissipation and relaxation.   

\subsection{Thermal state}\label{sec:c2therm} 
%I have to mention the toy model from Zoran here.  Did they give any fits for the early time?  Is their growth rate consistent with a universal behavior though?  I think it is.  
%
%
%
%width=8.6cm
\begin{figure}[h!]
\centering
\includegraphics[width=8.6cm]{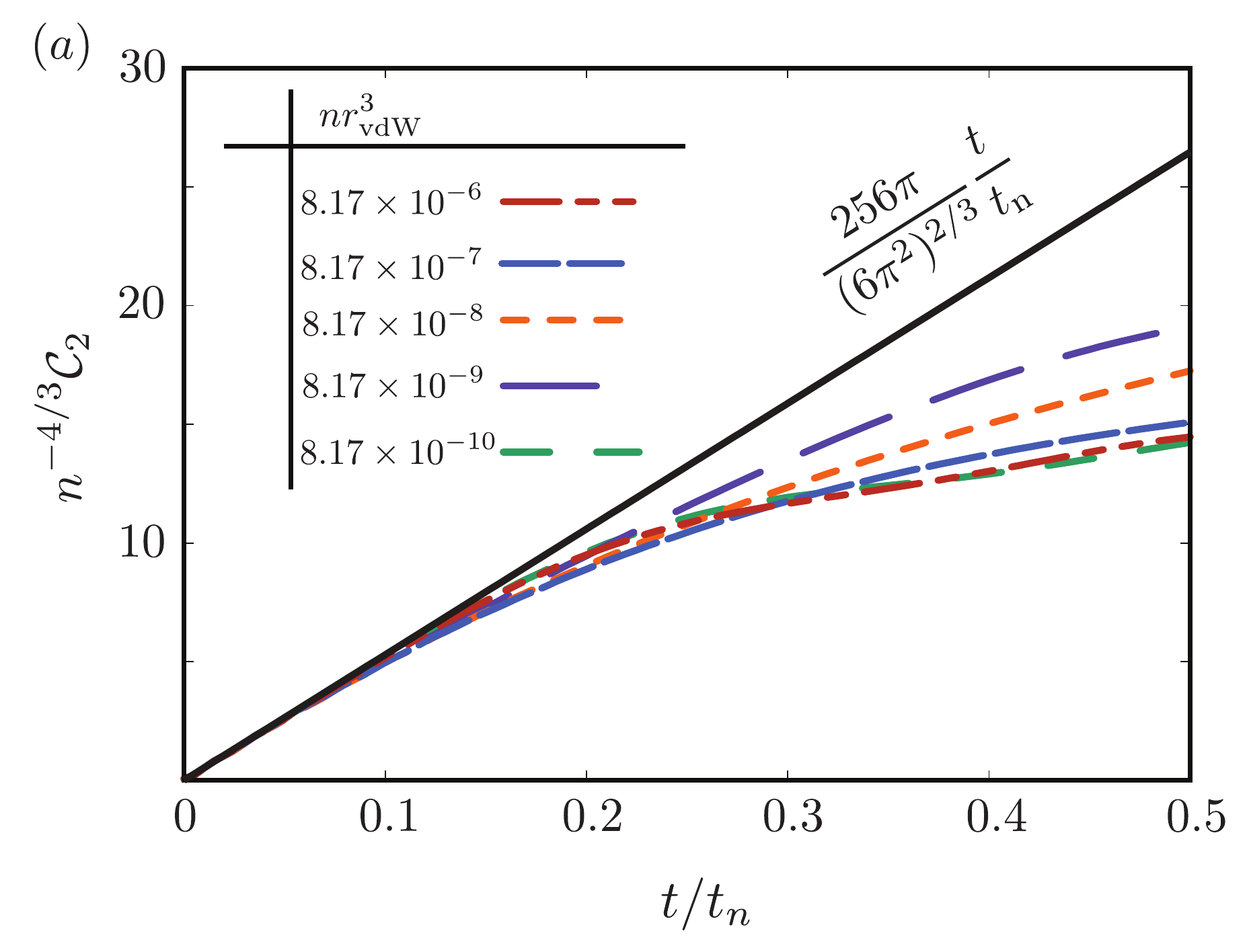}
\includegraphics[width=8.6cm]{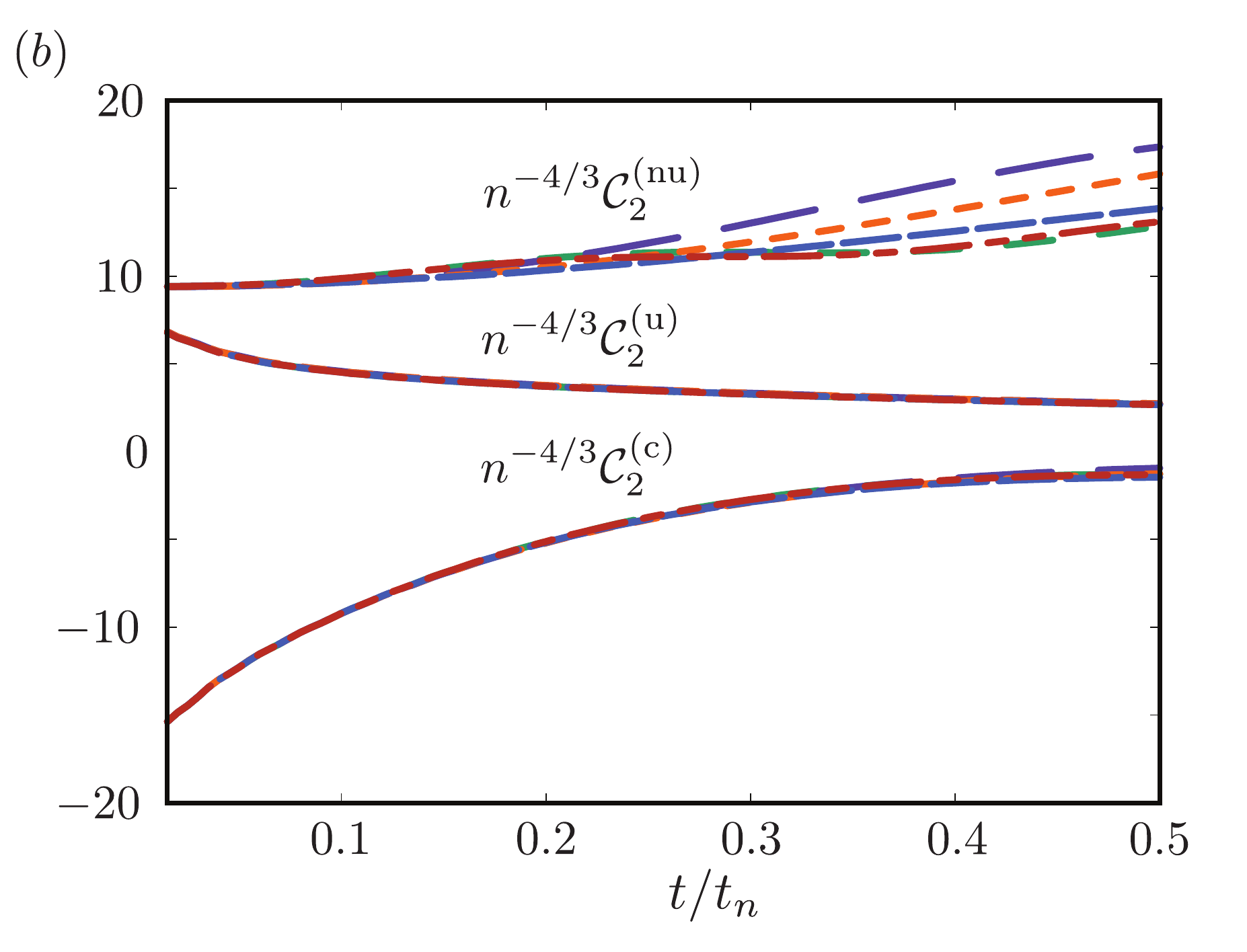}
\includegraphics[width=8.6cm]{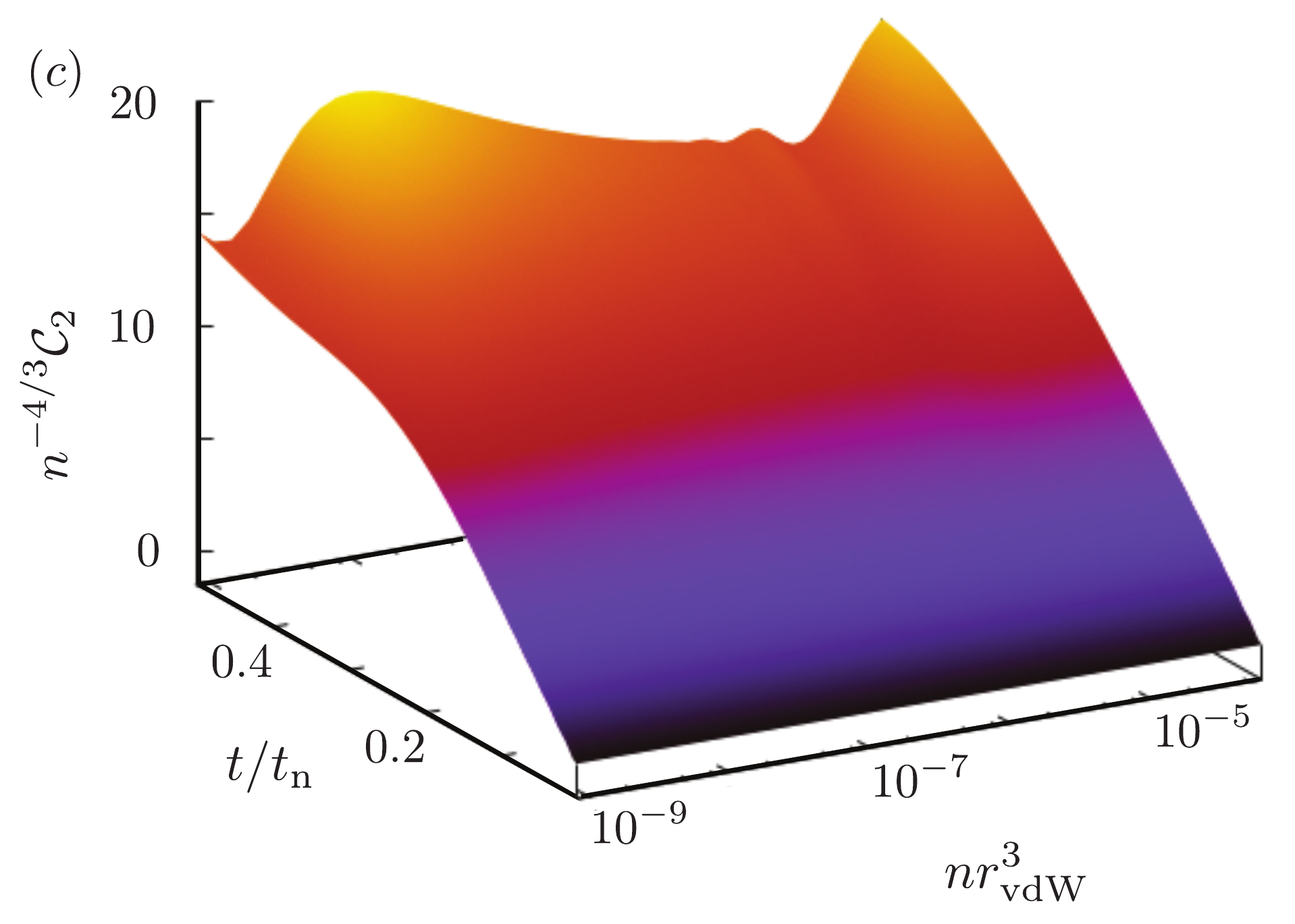}
\caption{(a) Dynamics of $\mathcal{C}_2(t)$ for thermal initial conditions with $\eta=0.06$ over a range of densities of experimental interest compared to the universal leading-order behavior [Eq.~\eqref{eq:c2limittherm}.]  (b) Evolution of the components of $\mathcal{C}_2$ over the same density range as used in (a).  (c) Dynamical surface showing the evolution of $\mathcal{C}_2$ over a range of densities.}\label{fig:c2therm}
\end{figure}
In this section, we present results for $\mathcal{C}_2(t)$ for an ideal thermal Bose gas quenched to unitarity by evaluating Eq.~\eqref{eq:c2step2} using the appropriate initial conditions given in Eq.~\eqref{eq:gzero}.  Although the equilibrium value for $\mathcal{C}_2$ was calculated in Refs.~\cite{PhysRevLett.112.110402,PhysRevLett.110.163202} as $\mathcal{C}_2^{eq}=32\pi\lambda^2n^2$, there are no analytic results in the literature for the post-quench growth of $\mathcal{C}_2(t)$ in this regime.  A two-body model was used in the supplementary materials of Ref.~\cite{Fletcher377} to estimate the time $\tau_2\approx0.1m\lambda^2/\hbar$ at which $\mathcal{C}_2(t)$ grows to $90\%$ of the model-specific equilibrium contact density.  However, this model cannot be expected to make quantitative predictions for a quenched thermal Bose gas because it both fails to satisfy the appropriate initial boundary conditions in Eq.~\eqref{eq:gzero} and is evaluated beyond  timescales $t_\lambda$, $t_\mathrm{n}$, and $t_s$ where a many-body treatment is necessary.  

Although we simulate a thermal state, there is no explicit temperature depedence in our model, and therefore we predict that the early-time contact dynamics are {\it temperature-independent} in the thermal regime.  This temperature-independent prediction is however valid only at times short relative to  $t_\lambda$, $t_\mathrm{n}$, and $t_s$, which is to say before individual atoms feel effects related to their neighbors and the finite coherence length in the problem $\propto\lambda$ \cite{cohen2011advances}.  Our results for the growth of $\mathcal{C}_2(t)$ in this regime are shown in Fig.~\ref{fig:c2therm}.  We show results for contact dynamics in the thermal regime out to $t\sim0.5t_\mathrm{n}$, with the caveat that in the high-temperature regime [$t_\lambda\gg t_\mathrm{n}$] this predictive range is further restricted.  We find that the leading-order growth is consistent with
\begin{equation}\label{eq:c2limittherm}
n^{-4/3}\mathcal{C}_2(t)=2!\frac{128\pi}{(6\pi^2)^{2/3}}\frac{t}{t_\mathrm{n}},
\end{equation} 
where the bunching factor $2!$ appears as a multiplicative correction to the BEC result in Eq.~\eqref{eq:c2limitcoherent}.  By taking just the leading-order dependence [Eq.~\eqref{eq:c2limittherm}], we make a crude estimate of $\tau_2$ by solving $\mathcal{C}_2(\tau_2)=0.9\mathcal{C}_2^{eq}$ to obtain $\tau_2=0.25 m\lambda^2/\hbar$.  For the temperature 370 nK of the gas used in Ref.~\cite{Fletcher377}, we find that $\tau_2\approx32$ $\mu$s, which is consistent with the conclusion that $\mathcal{C}_2$ saturates too quickly for its dynamics to have been resolved.  We note that failure to include the bunching factor $2!$ from Eq.~\eqref{eq:gzero} leads to a doubling of $\tau_2$, which would disagree with experimental findings.

As in Sec.~\ref{sec:c2coherent}, we find that the presence of the third boson is irrelevant at early times evidenced by agreement with Eq.~\eqref{eq:c2limittherm}, as shown in Fig.~\ref{fig:c2therm}(a), which can be obtained from a two-body model in the spirit of Ref.~\cite{PhysRevA.91.013616}.  However, the dynamics of $\mathcal{C}_2(t)$ depart from this initial growth behavior at $t\gtrsim0.1 t_\mathrm{n}$, which is even earlier than for the BEC case [see Fig.~\ref{fig:c2coherent}(a).]  This is the first indication of the different medium roles that may be played by the third boson depending on the initial state of the gas, and we return to this point shortly.  As in Sec.~\ref{sec:c2coherent}, the dynamics of $\mathcal{C}_2(t)$ become nonuniversal as the Efimov effect manifests through the third boson as log-periodicities and trimer beating both of which are  visible in the dynamical surface shown in Fig.~\ref{fig:c2therm}(c).  To study these effects, we return to the decomposition of $\mathcal{C}_2(t)$ in terms of universal, $\mathcal{C}_2^{(\mathrm{u})}$, non-universal, $\mathcal{C}_2^{(\mathrm{nu})}$, and coupled, $\mathcal{C}_2^{(\mathrm{c})}$, components whose behaviors are shown in Fig.~\ref{fig:c2therm}(b).  As for the BEC, both the log-periodicities and beating phenomenon arise from the contribution $\mathcal{C}_2^{(\mathrm{nu})}$ for $t\gtrsim0.1$.  A faint variation with the atomic density is also visible in $\mathcal{C}_2^{(\mathrm{c})}$.  Violation of the continuous scaling of $\mathcal{C}_2(t)$ is much more pronounced for the thermal gas on the order of $30\%$ by $t\sim0.5t_\mathrm{n}$.  Due to this variation, we find that the surface attains a maximum for densities satisfying 
\begin{equation}\label{eq:resthermc2}
k_\mathrm{n}R_\mathrm{3b}^{(j)}\approx 1.01, 
\end{equation}
which indicates that the coincidence of trimer size and interparticle spacing also enhances pair correlation growth in the thermal regime. 

\begin{figure}[t!]
\centering
\includegraphics[width=8.6cm]{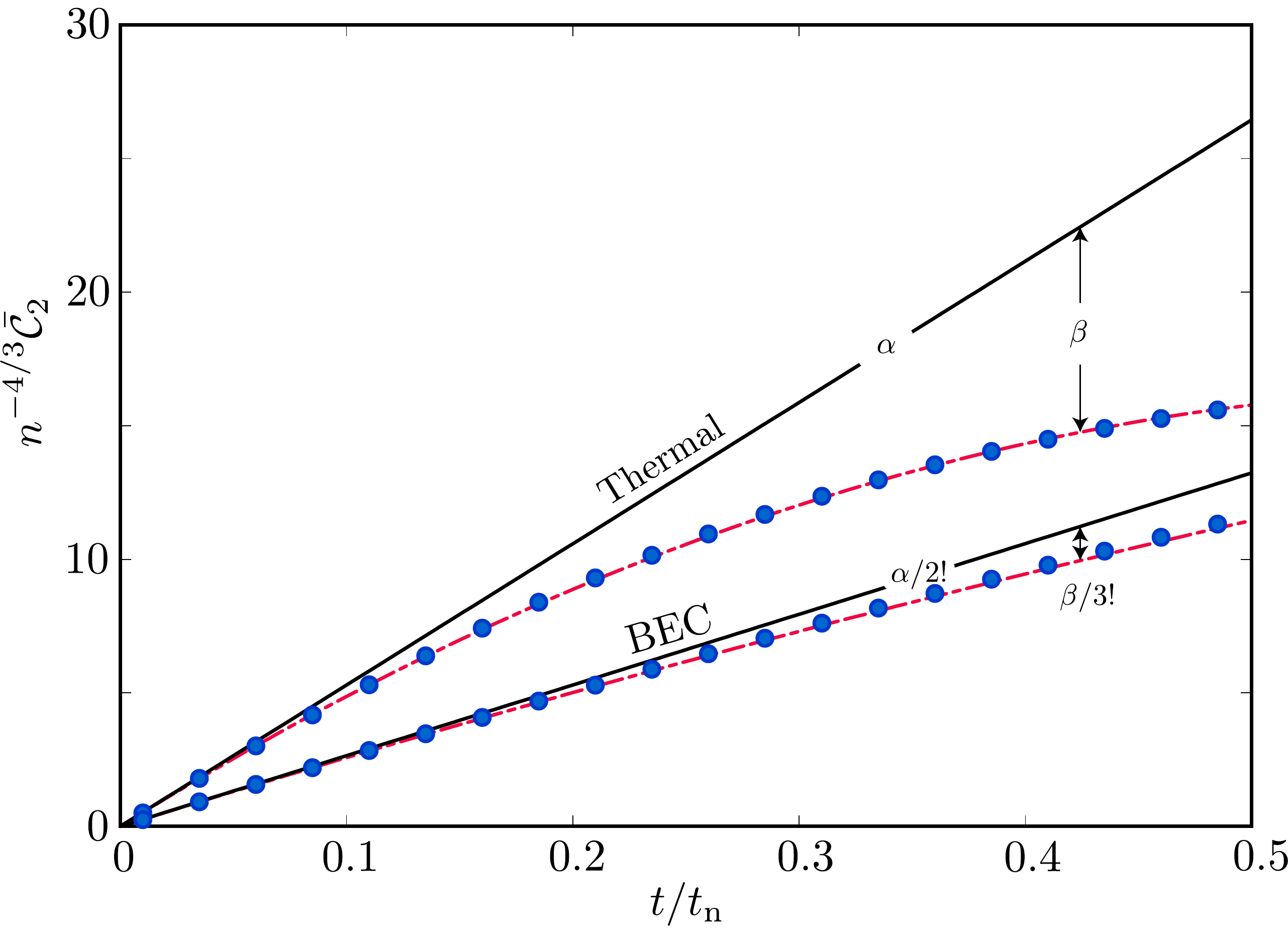}
\caption{Dynamics of $\bar{\mathcal{C}}_2(t)$ for both thermal and BEC initial conditions.  The solid black lines correspond to the leading order growth formulas.  The dot-dashed lines are the fits from Eqs.~\eqref{eq:avgfit}--\eqref{eq:avgfit2}.  The data points are from directly evaluating Eq.~\eqref{eq:c2step2}.\label{fig:c2avg}}
\end{figure}

By comparing the dynamics of $\mathcal{C}_2(t)$ for different initial initial conditions, the thermal or BEC nature of the third boson medium may be isolated in traces of atom-bunching effects.  Two-atom bunching effects are immediately clear by comparing the initial growth behaviors in Eqs.~\eqref{eq:c2limitcoherent} and \eqref{eq:c2limittherm}, where we the signature $2!$ is found.  Physically, this is a result of the primary conversion of bunched pairs into correlated two-body clusters with short-distance behavior $\propto r^{-1}$ in the sense of Sec.~\ref{sec:connections}.  Isolating the three-atom bunching signature $3!$ in the dynamics of $\mathcal{C}_2$ is however more subtle.  To investigate this, we average the dynamics of $\mathcal{C}_2(t)$ over a log-period in the atomic density 
\begin{equation}
\bar{\mathcal{C}}_2(t)=\frac{\int_{n_i}^{n_ie^{3\pi/s_0}} d\log(n)\mathcal{C}_2(t)}{3\pi/s_0},
\end{equation}
in order to isolate bunching effects from the Efimov effect.  We then fit the thermal data to 
\begin{equation}\label{eq:avgfit}
\bar{\mathcal{C}}_2(t)\approx\alpha\frac{t}{t_\mathrm{n}}+\beta\left(\frac{t}{t_\mathrm{n}}\right)^2\quad(\text{Thermal})
\end{equation}
and find that $\alpha=52.94\approx 256\pi/(6\pi^2)^{2/3}$ and $\beta=-7.12$ provides a reasonable fit as shown in Fig.~\ref{fig:c2avg}.  We test for the $3!$ signature by fitting the BEC data for $\bar{\mathcal{C}}_2(t)$ to
\begin{equation}\label{eq:avgfit2}
\bar{\mathcal{C}}_2(t)\approx\frac{\alpha}{2!}\frac{t}{t_\mathrm{n}}+\frac{\beta}{3!}\left(\frac{t}{t_\mathrm{n}}\right)^2\quad(\text{BEC}),
\end{equation}
which also provides a reasonable fit as shown in Fig.~\ref{fig:c2avg}.  The bunching factor $3!$ therefore makes the dominant contribution to the secondary dynamics of $\mathcal{C}_2(t)$ for the thermal Bose gas.  Physically, we interpret this as the secondary conversion of a third boson bunched in close proximity into a two-body cluster, which we conclude from Eq.~\eqref{eq:avgfit} is $3!$ more likely in the thermal case.  Conceptually, it is an intriguing question how this effect might cascade sequentially with other surrounding bunched bosons as the system evolves toward Fermi timescales.  This might be investigated theoretically, for instance, by using solutions of the unitary four-body problem \cite{PhysRevA.97.033621} to predict the contact dynamics through a straightforward extension of the procedure outlined in Sec.~\ref{sec:connections}.

\section{Post-quench dynamics of $\mathcal{C}_3(t)$.}\label{sec:c3}
%use this beginning section to derive the link?
In this section, we study the early-time dynamics of $\mathcal{C}_3(t)$ for BEC and thermal initial conditions, by enforcing Eq.~\eqref{eq:gzero} in the evaluation of Eq.~\eqref{eq:3bpsi} in Eq.~\eqref{eq:g3psi3}.  The dynamics of $\mathcal{C}_3(t)$ for a BEC quenched to unitarity were first studied in Ref.~\cite{PhysRevLett.120.100401}.  In this work, we revisit this study and extend it to the thermal Bose gas.  

We begin by deriving an expression for $\mathcal{C}_3(t)$ in terms of the three-body wave function in Eq.~\eqref{eq:3bpsi}.  As in Sec.~\ref{sec:c2}, details related to the specific basis of three-body eigenstates used can be found in App.~\ref{sec:appendix}.  From Eqs.~\eqref{eq:c3g3} and \eqref{eq:g3psi3}, we obtain 
\begin{equation}\label{eq:c3step1}
\frac{\mathcal{C}_3(t)4}{ns_0^2\sqrt{3}}\left|\Psi_{sc}(R,{\bf \Omega})\right|^2\underset{{R}\rightarrow 0}{=}\left|\sum_{s,j}c_{s,j}(t)\Psi_{s,j}(R,{\bf \Omega})\right|^2.
\end{equation}
As discussed in Sec.~\ref{sec:res3b}, only hyperradial eigenstates in the Efimov channel are nonzero in the limit $R\to0$, and therefore we ignore all universal channels in the above summation.  The hyperangular dependence of both sides of Eq.~\eqref{eq:c3step1} is identical and can be integrated over the solid angle $\int d{\bf \Omega}\equiv2\int d\alpha\sin^2 2\alpha\int_0^{\pi/2} d\hat{\boldsymbol{\rho}}\int d\hat{\bf r}$ to yield
\begin{equation}\label{eq:c3step2}
\left|\sum_{j}\frac{c_{s_0,j}(t)}{\sqrt{\langle F_j^{(s)}|F_j^{(s)}}\rangle}\times\frac{F_j^{(s)}(R)}{\sin\left[s_0\ln\frac{R}{R_t}\right]}\right|^2\underset{{R}\rightarrow 0}{=}\frac{12}{ns_0^2}\mathcal{C}_3(t).
\end{equation}
The above limit can now be taken without difficulty and, when using the trapped eigenstates of Sec.~\ref{sec:res3b}, can be calculated analytically along with the normalization factors $\langle F_j^{(s)}|F_{j}^{(s)}\rangle\equiv\int_0^\infty dR\ R |F_j^{(s)}(R)|^2$ as detailed in App.~\ref{sec:appendix}. 
% {\color{red} Did I put it in the Appendix?  See my old notes and code for the limiting expressions.  Make sure factors of aho squared update are correct now.}

\subsection{BEC}\label{sec:c3coherent}
%just put in the surfaces here.  It's not an exact repeat because I've plotted it in van der Waals units now.  
%
%
%
\begin{figure}[t!]
\centering
\includegraphics[width=8.6cm]{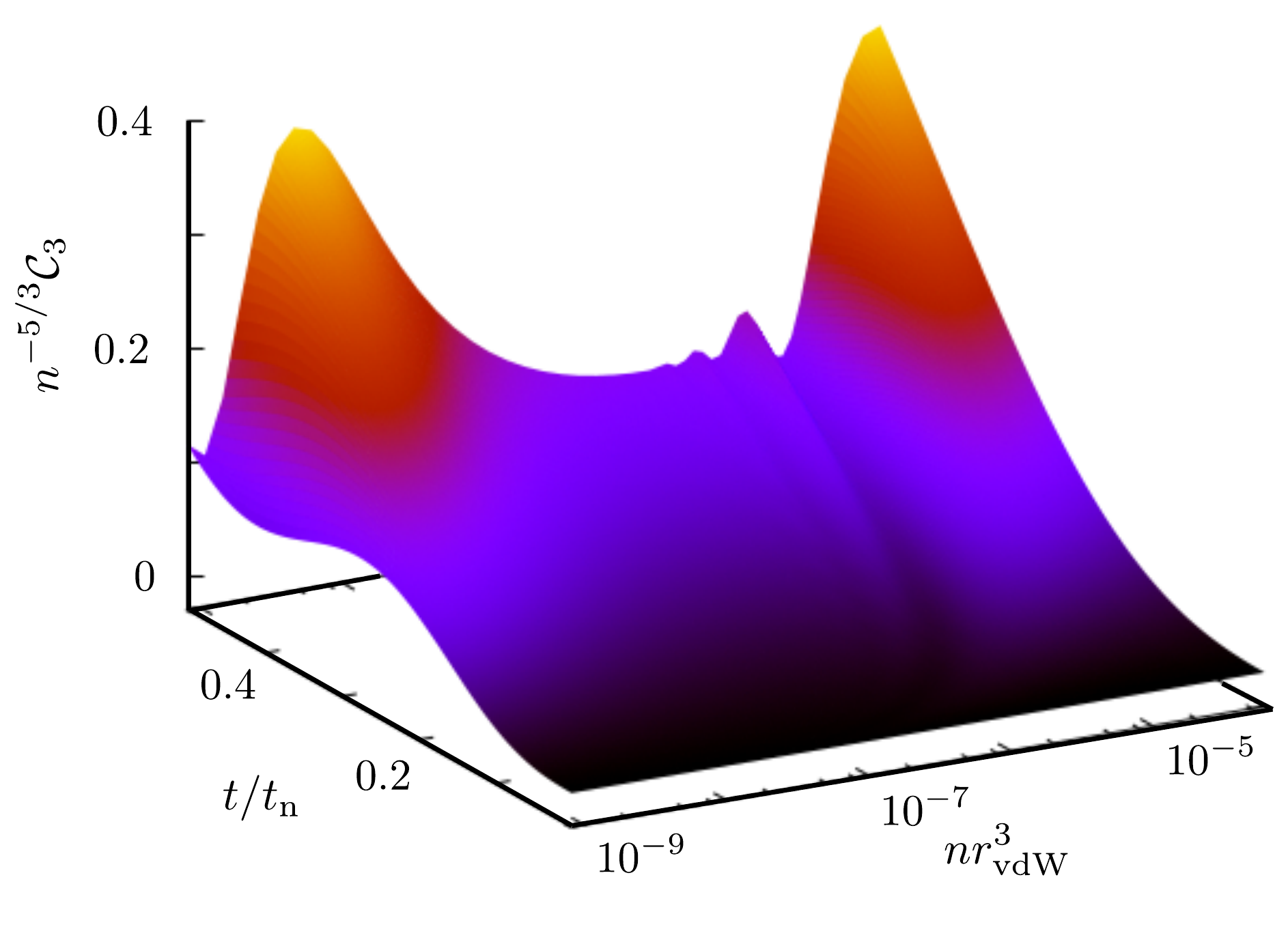}
\caption{Dynamical surface for the post-quench evolution of $\mathcal{C}_3(t)$ for BEC initial conditions with $\eta=0.06$ over a range of densities of experimental interest.   \label{fig:c3becsurface}}
\end{figure}
In this section, we review results for $\mathcal{C}_3(t)$ for a BEC quenched to unitarity first obtained in Ref.~\cite{PhysRevLett.120.100401}.  These results are revisited here both for reasons of completeness and to be contrasted against the results in Secs.~\ref{sec:c2coherent} and \ref{sec:c2therm} and for $\mathcal{C}_3$ for thermal initial conditions in Sec.~\ref{sec:c3thermal}.  The leading order growth of $\mathcal{C}_3(t)$ was fit in Ref.~\cite{PhysRevLett.120.100401} to 
\begin{equation}\label{eq:leadingc3}
n^{-5/3}\mathcal{C}_3(t)=0.55[1+3.09\times H(n,\kappa_*,t)]\left(\frac{t}{t_\mathrm{n}}\right)^2
\end{equation} 
 with unknown log-periodic function $H(n,\kappa_*,t)=H(ne^{3\pi j/s_0},\kappa_*,t)\in [0,1]$.  This log-periodic profile can be seen in the dynamical surface shown in Fig.~\ref{fig:c3becsurface}, which when plotted versus atomic density in van der Waals units applies broadly to atomic species satisfying $\eta\ll1$.  
 
 Although the visible beating phenomenon at the frequency of trimers at the scale of the interparticle spacing in Fig.~\ref{fig:c3becsurface} was first observed in Ref.~\cite{PhysRevLett.120.100401}, we now understand this to be a more general phenomenon in light of the results for $\mathcal{C}_2$ in Sec.~\ref{sec:c2}.  The visibility of these oscillations in time and of the log-periodic variations with the atomic density in the dynamical surface is however much greater for $\mathcal{C}_3$ than $\mathcal{C}_2$.  Intuitively, this agrees with the picture outlined in Sec.~\ref{sec:c2} that the Efimov effect is secondary in the dynamics of $\mathcal{C}_2$, entering only after the presence of the third boson is felt after a period of universal evolution.  Quantitatively, whereas the log-periodic oscillation of $\mathcal{C}_2$ was estimated in the $10-30\%$ range in Secs.~\ref{sec:c2coherent} and \ref{sec:c2therm}, it is the primary contribution for $\mathcal{C}_3(t)$, which is clear from Eq.~\eqref{eq:leadingc3}.  
 
 By inspecting the dynamical surface for $\mathcal{C}_3(t)$, we find that it attains a maximum for densities 
\begin{equation}\label{eq:resc3coherent}
k_\mathrm{n}R_\mathrm{3b}^{(j)}\approx 0.75,
\end{equation}   
which is within the error estimates $k_\mathrm{n}R_\mathrm{3b}^{(j)}\approx0.74(5)$ from Ref.~\cite{PhysRevLett.120.100401} obtained by comparing positions of the peaks at $t=0.5t_\mathrm{n}$ for two different forms of the initial three-body wave function.  We note that although the resonance conditions for $\mathcal{C}_2$ and $\mathcal{C}_3$ are slightly phase shifted, they both demonstrate the significance of scale-matching between Efimov trimer and interparticle spacing for few-body correlation growth in a BEC quenched to unitarity.

\subsection{Thermal state}\label{sec:c3thermal} 
%mention comparison's with Zoran's experiment.  
%we can now confirm that they grow slower than C2 in the thermal regime whereas it was just a qualitative analogy.  
%
\begin{figure}[t!]
\centering
\includegraphics[width=8.6cm]{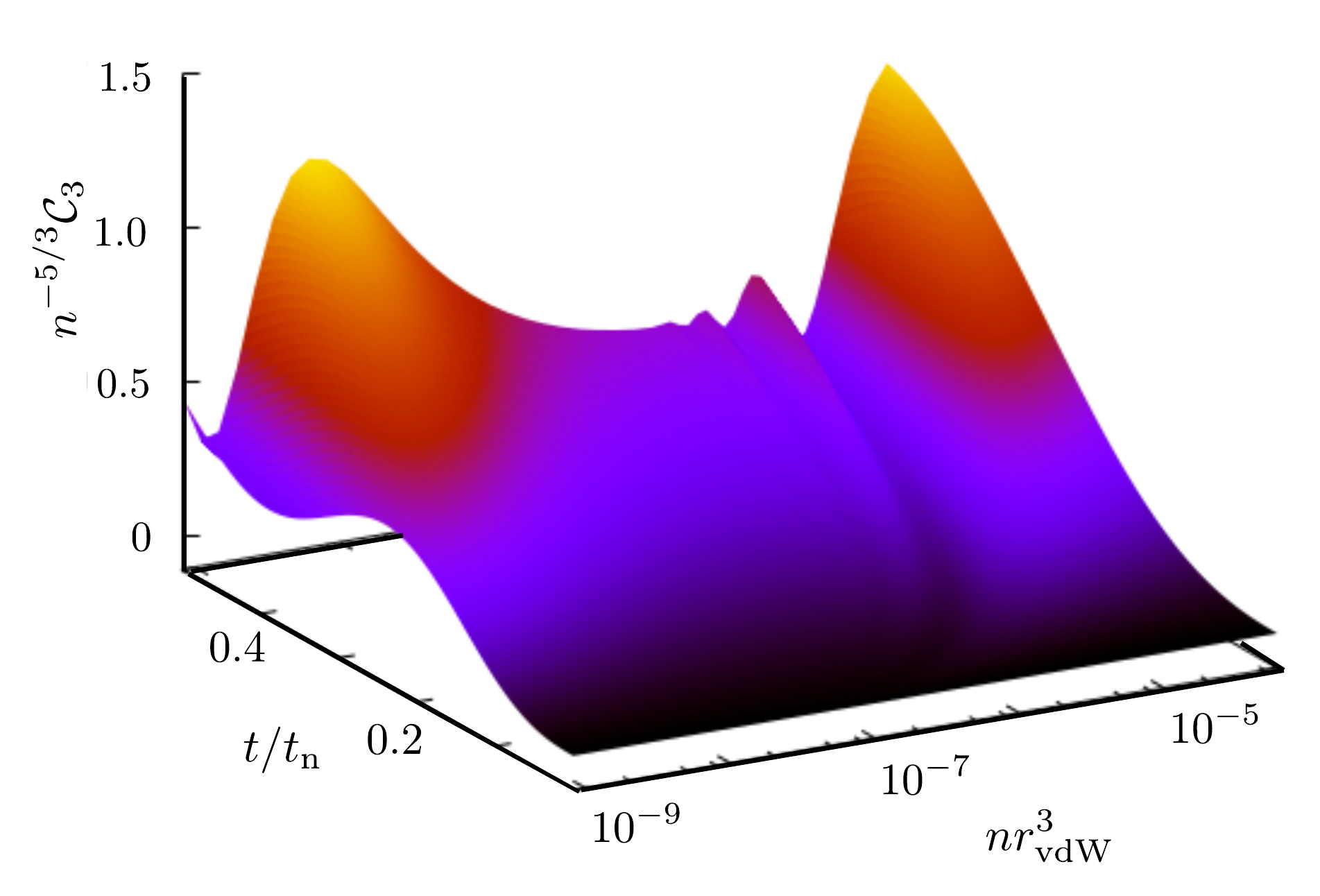}
\caption{ Dynamical surface for the post-quench evolution of $\mathcal{C}_3(t)$ for thermal initial conditions over a range of densities of experimental interest with $\eta=0.06$.  \label{fig:c3thermsurface}}
\end{figure}
In this section, we analyze the nonequilibrium dynamics of $\mathcal{C}_3(t)$ in the thermal regime.  From the dynamical surface shown in Fig.~\ref{fig:c3thermsurface}, we find that the early-time growth behavior of $\mathcal{C}_3(t)$ behaves as
\begin{align}\label{eq:c3leading2}
n^{-5/3}\mathcal{C}_3(t)=&3.17\left[1+2.91\times H \left(ne^{-0.1\pi/s_0},\kappa_*,t\right)\right]\nonumber\\
&\times\left(\frac{t}{t_\mathrm{n}}\right)^2,
\end{align} 
where the log-periodic profile function is well-approximated by a phase-shifted version of the function $H$ in Eq.~\eqref{eq:leadingc3}.  Comparing Eqs.~\eqref{eq:leadingc3} and \eqref{eq:c3leading2} reveals the three-body atom-bunching factor $3.17/0.55\sim3!$ in a ratio of the overall prefactors.  The log-periodic violations also reflect the three-body atom-bunching factor, which can be seen from the ratio $3.17\times2.91/1.70\sim3!$ of the prefactors of $H.$  

	The dynamics of $\mathcal{C}_3(t)$ have been studied experimentally in this regime in Ref.~\cite{Fletcher377}.  At the longest times studied in that work, $\mathcal{C}_3$ was found to approach the theoretical prediction $\mathcal{C}_3^{eq}\approx 3\sqrt{3}s_0\lambda^4n^3$ \cite{PhysRevLett.110.163202,PhysRevLett.112.110402}.  The early-time dynamics were however found to be consistent with zero.  Our temperature-independent prediction for the early-time dynamics of $\mathcal{C}_3(t)$ (Eq.~\eqref{eq:c3leading2}] holds for times $t<t_\lambda,t_\mathrm{n}$.  Therefore we consider the case $n\lambda^3=1$ where the latest times considered in our model [$t\sim0.5t_\mathrm{n}$] can be used as a prediction.   We find $\mathcal{C}_2(0.5t_\mathrm{n})/n^3\lambda^4\approx0.45$, which appears to be within the early-time experimental error bars in Ref.~\cite{Fletcher377}.  However, the time $t=0.5t_\mathrm{n}$ is less than the combined experimental duration of the RF pulse and shortest interrogation time, and therefore a direct quantitative comparison is not possible within our model.  We do however find qualitative agreement with the experimental finding that $\mathcal{C}_3$ grows slower than $\mathcal{C}_2$, which indicates a sequential buildup of clusters in the thermal regime.  

 By inspecting the dynamical surface for $\mathcal{C}_3(t)$, we find that it attains a maximum for densities 
\begin{equation}\label{eq:resc3therm}
k_\mathrm{n}R_\mathrm{3b}^{(j)}\approx 0.83,
\end{equation}   
which was used to estimate the phase shift of the log-periodic function $H$ in Eq.~\eqref{eq:c3leading2}.  Combined, the resonance conditions Eqs.~\eqref{eq:resc2coherent}, \eqref{eq:resthermc2}, \eqref{eq:resc3coherent}, and \eqref{eq:resc3therm} collectively indicate that the link between scale-matching between Efimov trimer and interparticle spacing and enhanced few-body correlation growth in a Bose gas quenched to unitarity is quite robust.

\section{Conclusion}\label{sec:conclusion}
In this work, we have analyzed the two- and three-body contact dynamics after quenching a Bose-condensed and thermal ultracold Bose gas to unitarity.  By connecting the correlation dynamics of this many-body system with solutions of the three-body problem, we search for signatures of the Efimov effect in the contacts.  We find that pair correlations are initially insensitive to three-body effects, evolving universally with the Fermi scales.  However, after a delay the medium effect of the third boson introduces intrinsic three-body effects, including log-periodicities and a trimer beating phenomenon.  Additionally, by comparing results for pair correlation growth  for thermal and BEC initial states, we find that the third boson carries a memory of the initial state of the medium.  For three-body correlations, we also find bunching signatures in the early-time dynamics of the thermal $\mathcal{C}_3$.  We find that log-periodicities and trimer beating first predicted in Ref.~\cite{PhysRevLett.120.100401} are robust, arising in the dynamics of both $\mathcal{C}_2$ and $\mathcal{C}_3$ for both thermal and BEC initial conditions.

In the thermal regime, our predicted contact dynamics are temperature-independent at times less than the thermal and Fermi times $t_\lambda$ and $t_\mathrm{n}$, respectively.  This constraint precludes direct quantitative comparison with the recent experimental results in Ref.~\cite{Fletcher377}.  However, our findings are qualitatively consistent with that work.  Namely, that $\mathcal{C}_2$ is saturated well-before the shortest interrogation times, and that $\mathcal{C}_3$ develops much more slowly in comparison. 

By extending the interferometric technique used in Ref.~\cite{Fletcher377} to the degenerate regime, the contact dynamics presented in this work and Ref.~\cite{PhysRevLett.120.100401} might also be tested.  Additionally, in the thermal regime where non-equilibrium many-body predictions at unitarity are lacking, the contact predictions presented in this work might be used as a benchmark.  This reasoning also applies in the degenerate regime where the pursuit of a many-body theory including the Efimov effect remains ongoing in the theoretical community \cite{KIRA2015185,PhysRevLett.89.210404,PhysRevA.98.051601}.  Finally,  we note that the method outlined in this work is completely general, and can therefore be extended straightforwardly to the scenario where channel couplings are important e.g. at finite scattering length \cite{Portegies2011}.  

\begin{acknowledgments}
{\it Acknowledgements.}  This work is supported by Netherlands Organisation for Scientific Research (NWO) under Grant 680-47-623.  J.P.D. acknowledges support from the U.S. National Science Foundation (NSF) under Grant No. PHY-1607204, and from the National Aeronautics and Space Administration (NASA).   We acknowledge discussions with Yuta Sekino and John Corson.   
\end{acknowledgments}

\appendix

\section{Further Details of the Unitary Trapped Three-Body Problem}\label{sec:appendix}
%careful that I remember the factor of 1/aho^2 difference between here and my notes.  see the pra notes.  
%like my PRL, there is no need to give the details of hte trapped solutions in the main text because they are not crucial to the analysis.  
An advantage of using analytic solutions to the trapped unitary three-body problem is that the calculation of the contact dynamics can be done fully analytically.  In this section, we begin by giving some details of the trapped three-body eigenstates from Ref.~\cite{PhysRevLett.97.150401,wernerthesis} for reasons of completeness \cite{typos}.  We then derive analytic results in Secs.~\ref{sec:overlaps}, \ref{sec:Atens} required to calculate the contact dynamics via Eqs.~\eqref{eq:c2step2}, \eqref{eq:c3step2}.  

Within each channel, the hyperradial eigenfunctions $F^{(s)}_j(R)$ satisfy the hyperradial Schr{\"o}dinger equation
\begin{equation}
\left[-\frac{\hbar^2}{2m}\left(\frac{d^2}{dR^2}+\frac{1}{R}\frac{d}{dR}\right)+U_s(R)\right]F_j^{(s)}(R)=E_\mathrm{3b}^{(j)}F^{(s)}_j(R)
\end{equation}
where $U_s(R)=\hbar^2s^2/2mR^2+m\omega^2R^2/2$ is the channel potential with trapping frequency, $\omega$, and associated trap length, $a_\mathrm{ho}=(\hbar/m\omega)^{1/2}$.  In the universal channels ($s^2>0$), the hyperradial eigenstates are given by
\begin{equation}
F_j^{(s)}(R)=e^{-R^2/2a_\mathrm{ho}^2}L_j^{(s)}(R^2/a_\mathrm{ho}^2)\left(\frac{R}{a_\mathrm{ho}}\right)^s,
\end{equation}
where $L_j^{(s)}$ is a generalized Laguerre polynomial of degree $j$.  The spectrum of three-body eigen-energies is given by $E_\mathrm{3b}^{(j)}=(s+1+2j)\hbar\omega$ where $j=0,1,2,\dots$  In the Efimov channel ($s=is_0$), they are given by
\begin{equation}
F_j^{(s_0)}(R)=\frac{a_\mathrm{ho}}{R}W_{E^{(j)}_\mathrm{3b}/2\hbar\omega,is_0/2}(R^2/a_\mathrm{ho}^2),
\end{equation}
where $W$ is a Whittaker function \cite{abramowitz1964handbook}.  The eigen-energy spectrum in the Efimov channel is obtained by solving
\begin{equation}
\arg\Gamma\left[\frac{1+is_0-E_\mathrm{3b}^{(j)}/\hbar\omega}{2}\right]+s_0\ln\frac{R_t}{a_\mathrm{ho}}=\arg\Gamma\left[1+is_0\right],
\end{equation}
which is understood $\mod \pi$.  We choose $R_t$ such that there is an Efimov trimer with binding energy $E^{(0)}_\mathrm{3b}=\hbar^2\kappa_*^2/m\approx 0.051\hbar^2/m r_\mathrm{vdW}^2$ in the free space limit $\omega\to0$. 

The normalization constant of the three-body eigenfunctions is given by
\begin{equation}
\mathcal{N}_{s,j}^2\equiv\left(\frac{2}{\sqrt{3}}\right)^3\frac{1}{\langle F_j^{(s)}|F_j^{(s)}\rangle\langle\phi_s|\phi_s\rangle},\label{eq:norm}
\end{equation}
where
\begin{align}
\langle \phi_s|\phi_s\rangle&\equiv\int d{\bf \Omega} |\phi_s({\bf \Omega})|^2.
\end{align}
 The components of $\mathcal{N}_{s,j}$ were calculated analytically in Ref.~\cite{wernerthesis} with result
 \begin{align}
 \langle \phi_s|\phi_s\rangle&=-\frac{12\pi}{s}\sin\left(\frac{s^*\pi}{2}\right)\nonumber\\
 \times&
\left[\cos\left(\frac{s\pi}{2}\right)-\frac{s\pi}{2}\sin\left(\frac{s\pi}{2}\right)-\frac{4\pi}{3\sqrt{3}}\cos\left(\frac{s\pi}{6}\right)\right],\\
 \langle F_j^{(s)}|F_j^{(s)}\rangle&=a_\mathrm{ho}^2\frac{\Gamma\left[s+1+j\right]}{2j!}\quad(s^2>0),\\
  \langle F_j^{(s_0)}|F_j^{(s_0)}\rangle&=a_\mathrm{ho}^2\frac{\pi\cdot \mathrm{Im} \psi\left(\frac{1-E_\mathrm{3b}^{(j)}/\hbar\omega+is_0}{2}\right)}{\sinh(s_0\pi)\cdot\left|\Gamma\left(\frac{1-E_\mathrm{3b}^{(j)}/\hbar\omega+is_0}{2}\right)\right|^2},
 \end{align}
 where $\psi$ is the digamma function.
 
 %do the full derivation here but maybe just give the main result and move the rest to the ``atoms'' paper. 
 %this is all correct and checked in teh code!  VEry good!
 We derive also expressions for the widths $\Gamma_j$ of each eigenstate in the Efimov channel using Eq.~\eqref{eq:widths}.  This requires that the extensive three-body contacts $C_3^{(j)}$ be obtained for each eigenstate.  We begin from the relation between the (non-normalized) triplet correlation function and the wave function for three bosons in vacuum \cite{pathria}
 \begin{equation}
 G^{(3)}({\bf r},{\bf r'},{\bf r''})=3!\Psi({\bf r},{\bf r'},{\bf r''}),
 \end{equation}
 where $|\Psi\rangle=|\Psi\rangle_{cm}|\Psi_{s_0,j}\rangle$ in terms of center of mass and relative wave functions.  Following Ref.~\cite{PhysRevA.86.053633}, we then integrate over the center of mass dependence, and take the $R\to0$ limit at fixed ${\bf \Omega}$ to relate with the extensive three-body contact
\begin{align}  
 3!\int d^3 C\ \Psi({\bf r},{\bf r'},{\bf r''})&=3!|\Psi_{s_0,j}(R,{\bf \Omega})|^2,\nonumber\\
 &\underset{{R}\rightarrow 0}{=}|\Psi_{sc}(R,{\bf \Omega})|^2\frac{8}{\sqrt{3}s_0^2}C_3.
 \end{align}
 The identical hyperangular dependence of $\Psi_{s_0,j}(R,{\bf \Omega})$ and $\Psi_{sc}(R,{\bf \Omega})$ allows us to integrate simply over $\int d{\bf \Omega}$ to obtain 
 \begin{equation}
\frac{s_0^2}{4\langle F_j^{(s_0)}|F_j^{(s_0)}\rangle}\left|\frac{F_j^{(s_0)}(R)}{\sin[s_0\ln R/R_t]}\right|^2\underset{{R}\rightarrow 0}{=}C_3^{(j)}.
\end{equation}
 From the asymptotic behavior of the Whittaker functions \cite{abramowitz1964handbook}, the above limit can be taken with the simple result
 \begin{align}
 a_\mathrm{ho}^2C_3^{(j)}&=\frac{s_0}{ \mathrm{Im} \psi\left(\frac{1-E_\mathrm{3b}^{(j)}/\hbar\omega+is_0}{2}\right)},\label{eq:c3trap}\\
  \frac{\Gamma_j}{\hbar\omega}&=4\eta\frac{1}{\mathrm{Im} \psi\left(\frac{1-E_\mathrm{3b}^{(j)}/\hbar\omega+is_0}{2}\right)},\label{eq:gammatrap}
 \end{align}
where $\Gamma_j$ is the width obtained via the relation given in Eq.~\eqref{eq:widths}. Equation~\eqref{eq:gammatrap} was first obtained in Ref.~\cite{wernerthesis}.  The free-space result $C_3=(e^{-2\pi/s_0})^n \kappa_*^2$ for the nth Efimov trimer can be obtained by taking asymptotic limits of the digamma function in Eq.~\eqref{eq:c3trap} \cite{atoms2018,abramowitz1964handbook}.  %UPDATE WITH ARXIV/ATOMS REF!!!!!!

%give the limiting behavior in the Efimov channel--see my old notes.  
%have to calculate analytic expressions for the widths.  
%perhaps cite the future atoms paper saying taht we have benchmarked against known results.  

%this should be in this paper.  
\subsection{Overlaps $c_{s,j}$}\label{sec:overlaps}
The overlaps for general $s$ are given by the integral 
\begin{equation}\label{eq:overlap}
c_{s,j}=\frac{3^{3/2}}{2^3} \int dR\ R^5 \int d\Omega\ \Psi_0(R,{\bf \Omega})\Psi_{s,j}^*(R,{\bf \Omega}).
\end{equation}
The hyperrangular integration can be easily performed and the total expression for the overlaps reduces to  \cite{wernerthesis}
\begin{align}
c_{s,j}&=\frac{\langle \phi_s|\phi_0\rangle}{\sqrt{\langle \phi_s|\phi_s\rangle\langle \phi_0|\phi_0\rangle}}\frac{\langle F_j^{(s)}|F_0\rangle}{\sqrt{\langle F_j^{(s_0)}|F_j^{(s)}\rangle \langle F_0|F_0\rangle}}, \label{eq:overlap1}\\
\langle \phi_s|\phi_0\rangle&=\frac{96\pi^{3/2}\sin\left[\frac{\pi s^*}{2}\right]}{4-(s^*)^2},\\
\langle \phi_0|\phi_0\rangle&=8\pi^3.
\end{align}
The mod-square of the overlaps obeys the following sum rule \cite{wernerthesis}
\begin{align}
|c_{s,j}|^2&=P(s)\frac{|\langle F_j^{(s)}|F_0\rangle|^2}{\langle F_j^{(s_0)}|F_j^{(s)}\rangle \langle F_0|F_0\rangle}, \label{eq:overlap2}\\
P(s)&\equiv\frac{|\langle \phi_s|\phi_0\rangle|^2}{\langle \phi_s|\phi_s\rangle\langle \phi_0|\phi_0\rangle}, \\
\sum_j |c_{s,j}|^2&=P(s),\\
\sum_{s}P(s)&=1,
\end{align}
where $|F_0\rangle$ is the hyperradial component of $|\Psi_0\rangle$
\begin{equation}
F_0(R)=\left(\frac{R}{a_\mathrm{ho}}\right)^2e^{-R^2/2B_1^2}\left[1-\left(\frac{R}{B_2}\right)^2\right],
\end{equation}
and 
\begin{equation}
\phi_0({\bf \Omega})=1.
\end{equation}
For general $s$, the total contribution to the norm of each channel is given by $\sum_{j}|c_{s,j}|^2=P(s)$ which was first obtained in Ref.~\cite{wernerthesis} as
\begin{eqnarray}\label{eq:overlapsdetail}
P(s)&=&\left[\frac{s\pi}{2}\sin\left(\frac{s\pi}{2}\right)-\cos\left(\frac{s\pi}{2}\right)+\frac{4}{3\sqrt{3}}\cos\left(\frac{s\pi}{6}\right)\right]^{-1}\nonumber\\
&&\times\frac{96 s\sin\left(\frac{s\pi}{2}\right)}{\pi(s^2-4)^2}.
\end{eqnarray}
We quote the analytic expression for $\langle F_j^{(s_0)}|F_0\rangle$ in Eq.~\eqref{eq:overlapsdetail}, which was first obtained in Ref.~\cite{PhysRevLett.120.100401} 
\begin{equation}\label{eq:overlapgen}
\langle F_j^{(s_0)}|F_0\rangle=\frac{1}{a_\mathrm{ho}}\left(\frac{1}{2}\ \Xi\left[\frac{1}{2},j\right]-\frac{1}{2B_2^2}\ \Xi\left[\frac{3}{2},j\right]\right),
\end{equation}
where
\begin{eqnarray}\label{eq:Xi}
\Xi[\alpha,j]&=&\frac{\left|\Gamma\left(\alpha+is_0/2+3/2\right)\right|^2}{\Gamma\left(\alpha+2-E_\mathrm{3b}^{(j)}/2\hbar\omega\right)}\times \left(\frac{1}{a_\mathrm{ho}^2}\right)^{(is_0+1)/2}\nonumber\\
&\times& \text{}_2F_1\left(a,\ b;\ c,\ z\right)\nonumber\\
&\times& \left(\frac{1}{2B_1^2}+\frac{1}{2 a_\mathrm{ho}^2}\right)^{-is_0/2-3/2-\alpha}.
\end{eqnarray}
The function $\text{}_2F_1$ is the Gauss hypergeometric function \cite{abramowitz1964handbook} with arguments
\begin{eqnarray}
a&=&3/2+\alpha+is_0/2,\\
b&=&\left(is_0-E_\mathrm{3b}^{(j)}/\hbar\omega+1\right)/2,\\
c&=&\alpha+2-E_\mathrm{3b}^{(j)}/2\hbar\omega,\\
z&=&\frac{1/B_1^2-1/a_\mathrm{ho}^2}{1/B_1^2+1/a_\mathrm{ho}^2}.
\end{eqnarray}
The overlaps $c_{s,j}$ for $s^2>0$ are also needed in the evalatuion of $\mathcal{C}_2(t)$ in Eq.~\eqref{eq:c2step2}.  We calculate analytic results for $\langle F_j^{(s)}|F_0\rangle$, and the relevant integrals can be found tabulated in Ref.~\cite{gradshteyn2007table}.   We obtain the result
\begin{equation}
\langle F_j^{(s)}|F_0\rangle=\frac{a_\mathrm{ho}^2}{2}\Upsilon[2,j,s]-\frac{a_\mathrm{ho}^4}{2B_2^2}\Upsilon[3,j,s],
\end{equation}
where 
\begin{align}
\Upsilon[\alpha,j,s]=&\frac{\Gamma\left[\frac{s}{2}+\alpha\right]\Gamma\left[s+j+1\right]}{j!\Gamma\left[s+1\right]}\left(\frac{1}{2}+\frac{a_\mathrm{ho}^2}{2B_1^2}\right)^{-s/2-\alpha}\nonumber\\
&\times\text{}_2F_1\left(-j,\frac{s}{2}+\alpha;s+1;\frac{1}{\frac{1}{2}+\frac{a_\mathrm{ho}^2}{2B_1^2}}\right).
\end{align}

%check all of this against what's in the code.  
\subsection{Evaluation of $A^{s,s'}_{j,j'}$}\label{sec:Atens}
To evaluate Eq.~\eqref{eq:c2step2} for the dynamics of $\mathcal{C}_2(t)$, integrals of the form 
\begin{equation}
A^{s,s'}_{j,j'}\equiv\sqrt{\frac{3}{2}}\int d z F^{(s)}_j(z)\left[F^{(s')}_{j'}(z)\right]^*
\end{equation}
must be evaluated, where we have defined the array $A^{s,s'}_{j,j'}$ as shorthand.  Below we obtain analytic expressions for $A^{s,s'}_{j,j'}$ for all relevant cases. 

%checked against code. 
{\bf Case I ($s^2>0,s'^2>0$)}:  For the universal channels, the array has the following integral form
\begin{equation}\label{eq:Auniversal}
A^{s,s'}_{j,j'}=\frac{1}{2}\sqrt{\frac{3}{2}}a_\mathrm{ho}\int dz \left(z\right)^{s/2+s'/2-1/2}e^{-z}L_j^{(s)}(z)L_{j'}^{(s')}(z).
\end{equation}
This integral can be evaluated analytically by expanding $L_{j'}^{(s')}(z)$ via the recurrence relation \cite{olver2010nist}
\begin{equation}\label{eq:laguerrerelation}
L_{j'}^{(s')}(z)=\sum_{l=0}^{j'}\begin{pmatrix}
(s'+s-1)/2+j'-l\\
j'-l\\
\end{pmatrix} L_l^{\left((s'-s-1)/2\right)}(z),
\end{equation}
where the $\left(\cdot\right)$ is the generalized binomial coefficient.  Equation~\eqref{eq:Auniversal} is now in a form which can be found tabulated in Ref.~\cite{gradshteyn2007table}, and we find with the help of symbolic mathematical software that
%checked and all good!
\begin{widetext}
\begin{align}
A^{s,s'}_{j,j'}&=\frac{1}{2}\sqrt{\frac{3}{2}}a_\mathrm{ho}(-1)^{j}\left(\frac{s+s'-1}{2}\right)!
\begin{pmatrix}
j'+\frac{s+s'-1}{2}\\
j'\\
\end{pmatrix}
\begin{pmatrix}
\frac{s'-s-1}{2}\\
j\\
\end{pmatrix}\nonumber\\
&\quad\quad\times
\text{}_3F_2\left(-j',-j-s,\frac{s'-s+1}{2};-j'-\frac{s+s'-1}{2},-j+\frac{s'-s+1}{2};1\right).\label{eq:Aunivs2}
\end{align}
\end{widetext}

%\begin{align}
%A^{s,s'}_{j,j'}&=\frac{1}{2}\sqrt{\frac{3}{2}}a_\mathrm{ho}\sum_{l=0}^{j'}(-1)^{j+l}\left(\frac{s+s'-1}{2}\right)!
%\begin{pmatrix}
%(s+j)\\
%l\\
%\end{pmatrix}\nonumber\\
%&\times
%\begin{pmatrix}
%(s'+s-1)/2+j'-l\\
%j'-l\\
%\end{pmatrix}
%\begin{pmatrix}
%(s'-s-1)/2+l\\
%j\\
%\end{pmatrix}.\label{eq:Aunivs2}
%\end{align}
%This result assumes $j>j'$, and in the case $j\leq j'$ a simple label interchange $s\leftrightarrow s'$, $j\leftrightarrow j'$ can be performed in Eq.~\eqref{eq:Aunivs2}.  

%checked against code
{\bf Case II ($s=s'=is_0$)}:  For the Efimov channels, the array has the following integral form
\begin{align}
A^{s_0,s_0}_{j,j'}&=\frac{1}{2}\sqrt{\frac{3}{2}}a_\mathrm{ho}\int dz \left(z\right)^{-3/2} W_{E^{(j)}_\mathrm{3b}/2\hbar\omega,is_0/2}\left(z\right)\nonumber\\
&\times W_{E^{(j')}_\mathrm{3b}/2\hbar\omega,is_0/2}\left(z\right).  
\end{align}
This integral can be found tabulated in Ref.~\cite{gradshteyn2007table}, and we obtain 
\begin{widetext}
\begin{align}
A^{s_0,s_0}_{j,j'}=\sqrt{\frac{3}{2}}a_\mathrm{ho}&\mathrm{Re}\left[\frac{\Gamma[1/2]\Gamma[1/2+is_0]\Gamma[-is_0]}{\Gamma\left[\frac{1-is_0-E^{(j')}_\mathrm{3b}/\hbar\omega}{2}\right]\Gamma\left[\frac{2+is_0-E^{(j)}_\mathrm{3b}/\hbar\omega}{2}\right]}\right]\nonumber\\
&\times \text{Re}\left[\text{}_3F_2\left(\frac{1}{2},\frac{1}{2}+is_0,\frac{1-E^{(j')}_\mathrm{3b}/\hbar\omega+is_0}{2};1+is_0,\frac{2-E^{(j)}_\mathrm{3b}/\hbar\omega+is_0}{2};1\right)                  \right],\label{eq:aefimov}
\end{align}
\end{widetext}
where $\text{}_3F_2$ is the generalized hypergeometric function \cite{olver2010nist}.  We note that for $j=j'$, Eq.~\ref{eq:aefimov} matches an expression first derived in Ref.~\cite{wernerthesis}.  The generalized hypergeometric function $\text{}_pF_q(a_1\dots a_p;b_1\dots b_q;1)$ is absolutely convergent on the unit circle if $\mathrm{Re}[\sum b_i-\sum a_j]>0$ \cite{olver2010nist}.  This works out to the requirement $E^{(j')}_\mathrm{3b}-E^{(j)}_\mathrm{3b}>0$.  If $E^{(j')}_\mathrm{3b}-E^{(j)}_\mathrm{3b}<0$, then the label transform $j\to j'$ performed on Eq.~\ref{eq:aefimov} will produce a convergent result.  

%checked against code
{\bf Case III ($s^2>0,s'=is_0$)}:  For the coupling of universal and Efimov channels, the array has the following integral form
\begin{align}
A^{s_0,s'}_{j,j'}&=\frac{a_\mathrm{ho}}{2}\sqrt{\frac{3}{2}}\int dz\ z^{s/2-1}e^{-z/2}L_j^{(s)}(z)\nonumber\\
&\times W_{E^{(j')}_\mathrm{3b}/2\hbar\omega,is_0/2}(z).\label{eq:dimentens}
\end{align}
This integral can be evaluated analytically by expanding the generalized Laguerre polynomial as \cite{olver2010nist}
\begin{equation}
L_j^{(s)}(z)=\sum_{l=0}^j (-1)^l\begin{pmatrix}
j+s\\
j-l\\
\end{pmatrix}
\frac{z^l}{l!}.
\end{equation}
Equation~\eqref{eq:dimentens} is now in a form that can be found tabulated in Ref.~\cite{gradshteyn2007table}, and we find 
\begin{align}
A^{s_0,s'}_{j,j'}&=\frac{a_\mathrm{ho}}{2}\sqrt{\frac{3}{2}}\sum_{l=0}^j \frac{(-1)^l}{l!} \begin{pmatrix}
j+s\\
j-l\\
\end{pmatrix}\nonumber\\
&\quad\quad\times\frac{\left|\Gamma\left[\frac{2l-is_0+s+1}{2}\right]\right|^2}{\Gamma[1-E^{(j')}_\mathrm{3b}/2\hbar\omega+l+s/2]}.\label{eq:efimovtensf}
\end{align}

%say that these formulas can be used to obtain C2 in a trap also which we do in a separate work. 

\section{Convergence}\label{sec:convergence}
In this section, we comment on the convergence of our results for the contact dynamics as a function of eigenbasis size.  For the dynamics of $\mathcal{C}_2$, the components ($\mathcal{C}_2^{(\mathrm{u})}$, $\mathcal{C}_2^{(\mathrm{nu})}$, $\mathcal{C}_2^{(\mathrm{c})}$) each have different convergence requirements.  It is therefore computationally more efficient to calculate each component separately.  The results presented in this work for $\mathcal{C}_2(t)$ were obtained using a basis consisting of the first 17 universal channels with 190 eigenstates per channel and 25 positive-energy eigenstates in the Efimov channel in addition to bound Efimov trimers overlapping significantly with the initial state.  For $\mathcal{C}_2^{(\mathrm{u})}$ and $\mathcal{C}_2^{(\mathrm{c})}$, we find convergence to 2-digits of precision beyond the decimal at $t=0$, which rapidly improves to 4-digits of precision or more by $t=0.5t_\mathrm{n}$.  For $\mathcal{C}_2^{(\mathrm{nu})}$, we find convergence to more than 4-digits of precision beyond the decimal at all times.  For the dynamics of $\mathcal{C}_3(t)$, the calculation is generally well converged at all times to at least 5 digits of precision beyond the decimal for using 100 eigenstate in the Efimov channel and the few bound Efimov trimers that overlap insignificantly with the initial state.

\bibliographystyle{apsrev4-1}
\bibliography{references}
\end{document}